\documentclass{aa}  

\usepackage{graphicx}
\usepackage{amsmath}
\usepackage{txfonts}
\usepackage{lipsum}
\usepackage{subcaption}                                         
\usepackage{lscape}                                
\usepackage{placeins}                                       
\usepackage[usenames]{color}
\usepackage{verbatim} 
\usepackage{siunitx}
\usepackage{booktabs}
\usepackage{natbib}

\begin{document}

   \title{A Jupiter-like radio aurora in the L3.5 dwarf LSPM J0036+1821}

   \author{D. Mart\'in-Carrero\inst{1}
        \and J.B. Climent \inst{1} \and P.A. Miles-P\'aez \inst{2} \and  J. Balm\'on-Ant\'on \inst{1}  \and J.C. Guirado \inst{1,}\inst{3} \and M.M. Kao \inst{4} \and C. O'Toole \inst{5} \and J.S.\ Pineda \inst{7} \and J.M. Vos \inst{5,6}}

   \institute{Departament d'Astronomia i Astrof\'isica, Universitat de Val\`encia, C. Dr. Moliner 50, E-46100 Burjassot, Val\`encia, Spain
    \and Centro de Astrobiolog\'ia, CSIC-INTA, Camino Bajo del Castillo s/n, 28692 Villanueva de la Ca\~nada, Madrid, Spain. \and Observatori Astron\`omic, Universitat de Val\`encia, Parc Cient\'ific, C. Catedr\'atico Jos\'e Beltr\'an 2, E-46980 Paterna, Val\`encia, Spain \and Lowell Observatory, 1400 W Mars Hill Road, Flagstaff, AZ 86001, USA \and School of Physics, Trinity College Dublin, The University of Dublin, Dublin 2, Ireland \and American Museum of Natural History, New York, NY 10024, USA \and University of Colorado Boulder, Laboratory for Atmospheric and Space Physics, 3665 Discovery Drive, Boulder, CO 80303,
USA}

   \date{Accepted October 6, 2026}
 
  \abstract
   {Ultracool dwarfs exhibit magnetically driven radio emission that reveals the presence of strong magnetic fields and complex magnetospheric processes. Their auroral emission shares similarities with that observed in magnetized Solar System planets, providing a framework to investigate their magnetic topology.}
   {We aim to characterize the magnetic activity, rotation, and atmospheric variability of the L3.5 dwarf LSPM J0036+1821 through a multiwavelength analysis.} 
   {We conducted a long-term monitoring campaign combining radio observations from the Very Large Array and very long baseline interferometry arrays with optical photometry from TESS, spanning 2019--2023. We analyzed time--series data to derive precise rotation periods and constructed phase--folded light curves and radio dynamic spectra. The radio lightcurves were modeled using an auroral framework analogous to Jupiter’s magnetosphere, incorporating both main oval and active field line components.}
   {We detect persistent, compact radio emission with significant variability and high circular polarization. Periodogram analysis yields a precise rotation period of 3.07941 $\pm$ 0.00020~h from radio data, consistent with the optical period derived from TESS (3.07908 $\pm$ 0.00038 h). The radio light curves display a stable morphology characterized by asymmetric, rotationally modulated pulses. Our modeling successfully reproduces the observed emission using a hybrid auroral scenario, with the main oval contributing a stable component and an active field line producing the dominant, variable bursts. Short-term variability is explained by small changes in the emission cone geometry or longitude of the active field line.}
   {The radio emission of LSPM J0036+1821 is consistent with auroral processes driven by the electron cyclotron maser instability in a structured magnetosphere. The observed behavior is consistent with a Jupiter-like auroral system, possibly involving companion-driven or internally-driven magnetospheric interactions analogous to those observed in the Solar System.}


   \maketitle
   \nolinenumbers

\section{Introduction}
\label{sec:Intro}

Ultracool dwarfs are objects with spectral types later than M7, which includes very low-mass stars, brown dwarfs, and planetary-mass objects with effective temperatures < 2700 K. \citep{KP1999, MT1999}. These low temperatures lead to the formation of condensates, which can aggregate to form cloud-like structures \citep{1996A&A...308L..29T}. These are likely heterogeneously distributed across the atmosphere of the dwarf \citep{2009ApJ...701.1534A,2014Natur.505..654C}, leading to photometric and spectroscopic variations modulated by rotation \citep{2014ApJ...793...75R,2021AJ....161..224T}.

The serendipitous discovery of radio emission in these objects \citep{2001Natur.410..338B} demonstrated the presence of strong magnetic fields at the level of a few kG in these largely neutral atmospheres. Subsequent studies revealed that around 15 -- 20$\%$ of ultracool dwarfs are radio-loud \citep{2024MNRAS.527.6835K}. Their radio emission typically consists of two different components: (i) a quiescent component, generally attributed to gyro or synchrotron emission produced by either magnetic reconnections or by the presence of a radiation belt around the dwarf \citep{2023Sci...381.1120C, 2023Natur.619..272K} and (ii) a pulsed, highly polarized and rotationally modulated bursty emission attributed to the Electron Cyclotron Maser Instability \citep[ECMI;][]{2008ApJ...684..644H}. 
This auroral radio emission is analogous to that observed in magnetized planets of the Solar System, such as Jupiter and Saturn \citep[e.g.,][]{2004AdSpR..33.2045Z,2011JGRA..116.4212L, 2017A&A...604A..17M}. The close distance to these planets allows a detailed characterization of their auroral emitting regions, revealing a complex structure. In particular, auroral radio emission can be driven by the interaction between the planet and its moons, as well as by magnetospheric currents associated with a co-rotating plasma disk. The physical similarities between ultracool dwarfs and these giant gas planets suggest that comparable mechanisms may operate in both regimes \citep{2008ApJ...684..644H, 2012ApJ...760...59N,2017MNRAS.470.4274T}. In this context, modeling the auroral radio emission of ultracool dwarfs using analogies with the well-characterized magnetospheres of Solar System planets provides a valuable tool to constrain their geometry, properties that are otherwise extremely difficult to determine observationally \citep{2024A&A...682A.170B, 2024A&A...692A..66K, 2025ApJ...987....7G}.

LSPM J0036+1821 is an L3.5 dwarf \citep{2000AJ....120.1085G, 2000AJ....120..447K} located at a distance of $8.7357 \pm 0.0001$~pc \citep{2020yCat.1350....0G}. Radio emission from this object was first detected by \cite{2002ApJ...572..503B}, who reported both quiescent and auroral emission. \cite{2008ApJ...684..644H} monitored this object during $\sim$10~h using the Very Large Array (VLA) in C-band (4 - 6~GHz), detecting highly circularly polarized auroral bursts modulated with a period of $3.08 \pm 0.05$~h, attributed to the rotation period of the dwarf. Subsequent searches for radio emission at lower frequencies with the VLA at 325 MHz and 1.4~GHz resulted in non-detections \citep{2011AJ....142..189J}. \cite{2017MNRAS.465.1995M} observed the target simultaneously in S (2-4~GHz) and C (4-8~GHz) bands; these authors detected only quiescent emission, with no evidence of pulses, although their observations did not cover a full rotation period of the object, which may explain the absence of detected bursts.
In addition to its radio activity, LSPM J0036+1821 has been reported to exhibit photometric and spectroscopic variability at optical and infrared wavelengths \cite[e.g.,][]{2007ApJ...668L.163L, 2013MNRAS.428.2824K, 2013ApJ...779..101H, 2015ApJ...799..154M, 2026A&A...710A..22O}, with a rotation period of $\sim$3 h. \cite{2010ApJ...715..724B} and \cite{2013ApJ...767..110P} reported the possible presence of an L5--6 companion at a projected separation of $89.5 \pm 11.4$ mas, corresponding to a physical separation of $0.78 \pm 0.10$ AU. However, this putative companion has not been confirmed yet.

In this work, we present the results of a multiwavelength campaign aimed to characterize the rotation and radio emission
of LSPM J0036+1821.

\section{Observations and Data Reduction}
\label{sec:Obs_DR}
Below, we describe VLA (Karl G. Jansky Very Large Array) observations, VLBI (Very Long Baseline Interferometry) observations obtained with the EVN (European VLBI Network) and the VLBA (Very Long Baseline Array), and TESS (Transiting Exoplanet Survey Satellite) observations of LSPM J0036+1821, spanning 2019--2023. A summary of these is listed in Table ~\ref{tab:Obs}.

\subsection{VLA observations}
\label{subsec:VLA}

\begin{figure}[h!]
        \centering
        \includegraphics[width=0.93\columnwidth]{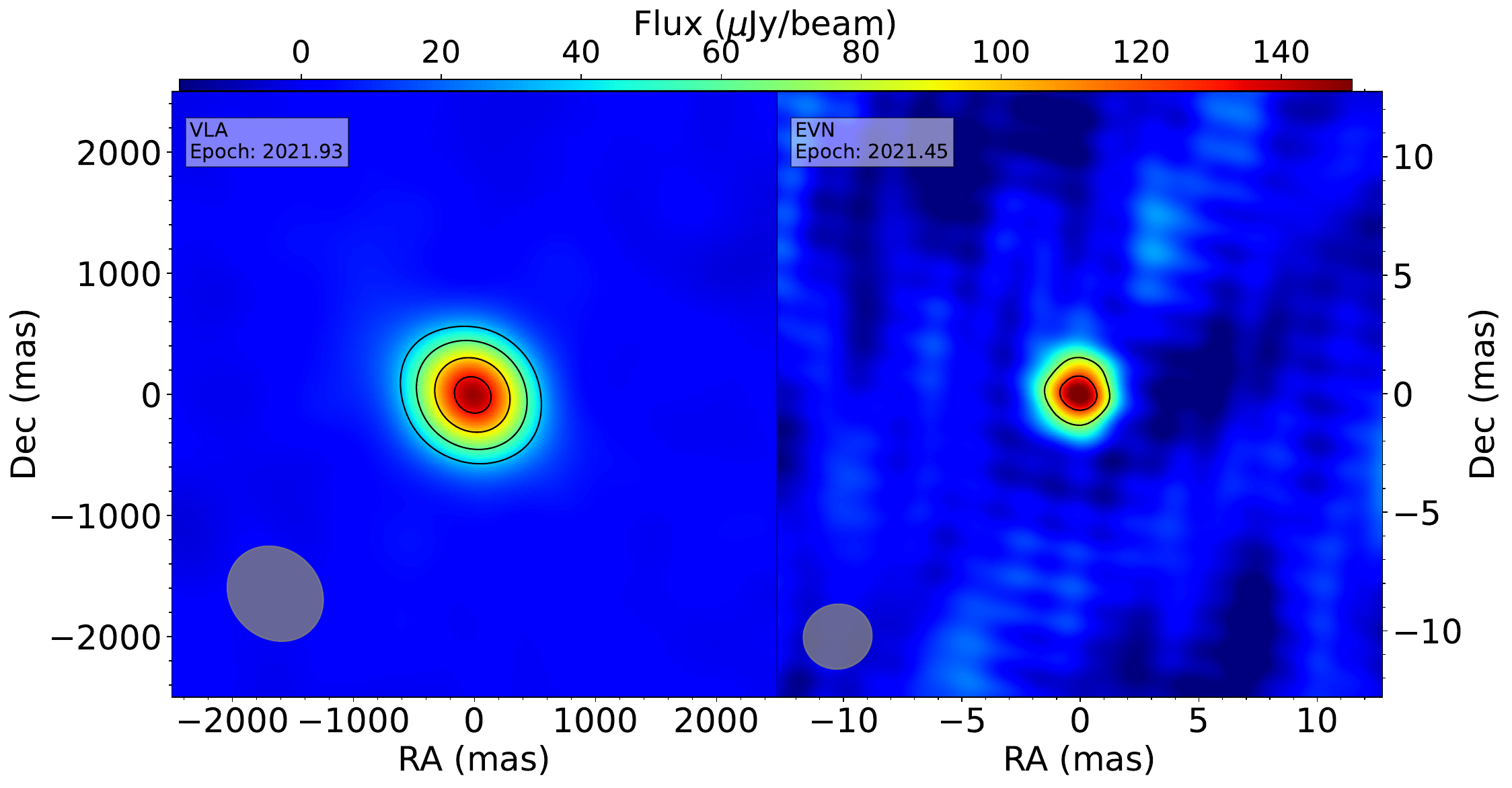}
        \caption{Reconstructed Stokes I images centered at the expected coordinates of LSPM J0036+1821. We show the maps corresponding to the VLA (left) and a representative epoch of the EVN (right). The black contours represent the 3$\sigma$, 5$\sigma$, 8$\sigma$ and 11$\sigma$ detection levels. The gray ellipses represented in the left lower part of each reconstructed image represents the beam size (see Table ~\ref{tab:Obs}).}
        \label{fig:Radio_img}%
    \end{figure}

LSPM J0036+1821 was observed in X-band (8 - 12~GHz) with the VLA on the night of 2021 December 7 (epoch 2021.93, project VLA/SR0228, PI: J, Vos). The array was in B-configuration, and observations were conducted in phase-referencing mode, using 27 antennas. The nearby blazar J0042+2320, which is included in the International Celestial Reference Frame (ICRF), was used as a phase calibrator, following a 10-minute duty cycle consisting on 8~min on target and 2~min on calibrator, for a total observing time of 8.7 h. Additionally, the standard VLA flux and bandpass calibrator 3C48 was observed at the beginning and end of the session. The integration time was 3~s, and the total bandwidth was divided into 32 spectral windows. The total integration time on target was 392~min, corresponding to approximately 75\% of the session. Data calibration was performed using the standard VLA pipeline and verified through inspection of the NRAO observing logs. Right Circular Polarization (RCP) and Left Circular Polarization (LCP) time-averaged images were produced using the \texttt{clean} task within the Common Astronomy Software Applications (CASA) package \citep{2007ASPC..376..127M} using a 200x200 pixel image size, a cell size of 0.05 arcsec and natural weighting. We searched for emission at the expected position of the target after proper-motion propagation; we detected a compact, point-like source with a total flux density of 160 $\pm$ 4~$\mu$Jy (see Figure ~\ref{fig:Radio_img}). Inspection of the field revealed no additional sources within the primary beam. In order to directly compare VLA and VLBI data, we exported the calibrated data from CASA using the \texttt{exportuvfits} task and imported into the NRAO Astronomical Image Processing System \citep[AIPS;][]{1996ASPC..101...37V} using the task \texttt{fitld}. Separate time-series of flux density values at RCP, LCP, Stokes I and Stokes V were created using the task \texttt{dftpl} averaging for all baselines and spectral windows with a time resolution of 60\,s.

The high sensitivity of the VLA data enabled the production of wideband dynamic spectra, which we outline in turn: (i) we verified that no additional sources within the primary beam could contaminate the target's signal; (ii) we introduced a phase delay in the visibilities using the CASA routine \texttt{fixvis} to place our target in the phase-center and ensure that the target emission appears in the real component of the visibilities; (iii) we used the CASA \texttt{plotms} routine to average the RCP and LCP visibilities across all baselines and channels at a temporal and spectral resolution of 60\,s and 128\,MHz, respectively; and (iv) Stokes I and V dynamic spectra were also produced from the calibrated RCP and LCP correlated amplitudes (RR and LL, respectively), with $I = \frac{1}{2} (RCP + LCP)$ and $V = \frac{1}{2} (RCP - LCP)$.

\subsection{VLBI observations}
\label{subsec:VLBI}
    
EVN observations were performed in C-band (5~GHz) in 2019 and 2020 (codes EC067C, EC077C, PI: J, Climent). The observations were performed using phase-referencing mode. We used the blazar J0038+1856 as the phase calibrator for both sessions. The total bandwidth was divided into 8 spectral windows each with a bandwidth of 16 MHz. For the first session (2019.17), the duty cycle was 306~s, consisting of 68~s on calibrator and 238~s on source. For the second session (2021.45), the duty cycle was 366~s, with 118~s on the calibrator and 248~s on the target. These setups resulted in a total on-source integration time of 1.65~h (66\%   of the session) and 3.85~h (75\%   of the session) for 2019.17 and 2021.45 epochs, respectively. 
The data was calibrated using the NRAO AIPS (Astronomical Image Processing System) software following standard procedures implemented in the VLBARUN pipeline: (i) we applied a-priori amplitude corrections; (ii) we applied ionospheric corrections based on total electron content maps; (iii) we updated the Earth Orientation Parameters (EOP); (iv) we performed digital sampler corrections; (v) we removed phase-jumps between IFs using a fringe fitting on the phase calibrator; (vi) we transfer the local fringe fitting solution to the target; (vii) we performed a global fringe on the calibrator over the full observation; and (viii) we finally interpolate this global fringe solution to the targets visibilities. 
We also included VLBA archival data in our analysis. These observations were conducted in C-band (5~GHz) during 2020 and 2021 (project codes BC250 and BC267). The observations were performed using phase-referencing mode, in same way as the EVN observations but using the blazar J0028+2000 as the phase calibrator. The duty cycle consisted of 46~s on the calibrator and 106~s on the target, yielding a total cycle time of 152 s. Each observing epoch lasted $\sim$3~h, corresponding to an on-source integration time of $\sim$1.5 h. The data were divided into 4 spectral windows of 128 MHz each. VLBA data reduction was performed following the same procedure as the one described for the EVN data.

For both VLBI datasets (VLBA and EVN observations), we produced naturally weighted images using Difmap (Shepherd 1997). In all epochs, we detected compact emission at the proper-motion corrected position of LSPM J0036+1821. We measured the positional offset of the source relative to the phase center using Difmap, and subsequently generated the corresponding time series with the \texttt{dftpl} task in AIPS, averaging all spectral channels and baselines every $\sim$120 s. The exact temporal binning varies slightly between epochs to account for differences in the rms noise of the observations.
\subsection{TESS observations}
\label{subsec:TESS}

LSPM J0036+1821 was observed by TESS \citep[][]{2014SPIE.9143E..20R} during Sectors 17 and 57, both acquired with a 2-minute cadence. For our analysis, we used the pipeline-calibrated data products released by TESS. Specifically, we used the Pre-search Data Conditioned Simple Aperture Photometry (PDCSAP) in both sectors, which are already corrected for common instrumental systematics.

\section{Results}
\label{sec:Res}
\subsection{Radio imaging detections}
\label{subsec:Radio_img}

We detect compact, point-like emission at the expected position of LSPM J0036+1821 at all epochs and instruments. A complete summary of the peak flux densities for each epoch is provided in Table ~\ref{tab:Obs}. Significant variability is observed between epochs, with the peak flux density ranging from 78~$\mu$Jy to 307~$\mu$Jy, encompassing the flux values reported by \cite{2008ApJ...684..644H} from VLA observations. The origin of this epoch-to-epoch variability is discussed in Section ~\ref{subsec:LC_tempevol}. LCP (negative Stokes V) dominates in most epochs with values ranging between -154 and -55~$\mu$Jy. Only two epochs, BC267B1 (2021.12) and EC067C (2019.17), show overall positive Stokes V detections of 50 and 60~$\mu$Jy, respectively. A detailed characterization of the polarization properties and their temporal evolution is presented in Section \ref{subsec:Multi_LC}.

The milliarcsecond angular resolution provided by the VLBI observations (EVN and VLBA datasets) allows us to search for spatially resolved emission at the position of the target. In particular, we investigate the presence of extended or structured emission similar to that reported by \cite{2023Sci...381.1120C} and \cite{2023Natur.619..272K} in the M8.5 dwarf LSR J1835+3259. In all epochs, the source remains unresolved and consistent with a point-like morphology. Using the smallest synthesized beam  (epoch 2021.45; see Table ~\ref{tab:Obs}) and the Gaia distance, we derive an upper limit on the size of the emitting region of $\sim$0.025~AU ($\sim$52~$\rm R_{Jup}$). The reconstructed radio images are shown in Figure ~\ref{fig:Radio_img}, where we present one representative image at VLBI scales, as all epochs display a similar compact appearance.

\subsection{Radio periodicity}
\label{subsec:Periodic_Rad}
\begin{figure}[h!]
   \centering
   \includegraphics[width=0.89\columnwidth]{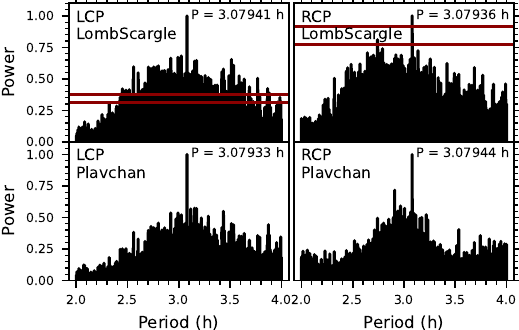}
      \caption{Periodogram analysis of the radio lightcurves of LSPM J0036+1821. Top panels show the LS periodogram for the LCP (left) and RCP (right) data. Horizontal red lines show the 0.1\%  and 1\%  False Alarm Probability levels. Bottom panels show the Plavchan periodogram using an outlier fraction of 10\%  and a phase smoothing width of 0.1.}
         \label{fig:Radio_period}
   \end{figure}

To determine the radio rotation period of LSPM J0036+1821 we analyzed the VLBI and VLA datasets together (lightcurves are shown in Figure ~\ref{fig:radio_lc} and Figure ~\ref{fig:VLA_lc_ev}). Since the large-scale magnetosphere is assumed to rotate as a rigid body \citep{2018Natur.555..227G}, both the RCP and LCP components should share the same underlying rotation period. We therefore first applied a Lomb-Scargle \citep[LS;][]{1976Ap&SS..39..447L, 1982ApJ...263..835S} periodogram independently to the RCP and LCP lightcurves. A prominent peak is identified for each component, consistent with previous measurements, at P = $3.07936$~h and P = $3.07941$~h for LCP and RCP, respectively (Figure ~\ref{fig:Radio_period}). Varying the frequency sampling of the LS periodogram does not produce significant changes in the derived periods. Even if the Lomb-Scargle periodogram is optimized to detect sinusoidal signals, it is still able to detect periodic signals in radio aurorae which are not necessarily sinusoidal.

To independently verify these results, and following the work done by \cite{2018ApJS..237...25K}, we also employed the Plavchan periodogram \citep{2008ApJS..175..191P}, a binless variation of the Phase Dispersion~minimization technique \citep[PDM;][]{1978ApJ...224..953S}. This method is independent of the pulse shape and is therefore better suited for detecting non-sinusoidal periodic signals. Figure ~\ref{fig:Radio_period} shows the resulting periodograms. 
To estimate the uncertainties, we performed a bootstrap analysis by simulating 20 RCP and LCP lightcurves. In each simulation, every data point was randomly perturbed according to a Gaussian distribution with a standard deviation equal to its measured $1 \sigma$ uncertainty. For each synthetic lightcurve, we computed the Plavchan periodogram and recorded the period corresponding to the highest-power peak. This procedure yields periods of P = $3.07948 \pm 0.00015$~h and P = $3.07934 \pm 0.00012$~h for the RCP and LCP components, respectively. We therefore adopt a final radio rotation period of P = $3.07941 \pm 0.00020 $~h for LSPM J0036+1821. The central value corresponds to the average of the periods derived from the RCP and LCP data, while the quoted uncertainty is obtained from the propagated errors of the two individual measurements.

\subsection{Optical periodicity}
\label{subsec:Periodic_Opt}

We searched for rotationally modulated emission in the TESS PDCSAP data for sectors S17 and S57. We model the lightcurves using a similar approach to that used in \cite{2021A&A...651L...7M} and \cite{2023MNRAS.521..952M} that we describe briefly here; we decompose the observed signal into two components, (i) a low-frequency term accounting for residual instrumental systematics and long-term trends not fully removed by the TESS pipeline \citep{2019AAS...23320201V} which is modeled using a Gaussian Process (GP) with a Mat\'ern-3/2 kernel \citep{2006gpml.book.....R} and (ii) a periodic component associated with the intrinsic emission of the target modeled as a sine function. 

   \begin{figure}[h!]
            \centering
            \includegraphics[width=0.876\columnwidth]{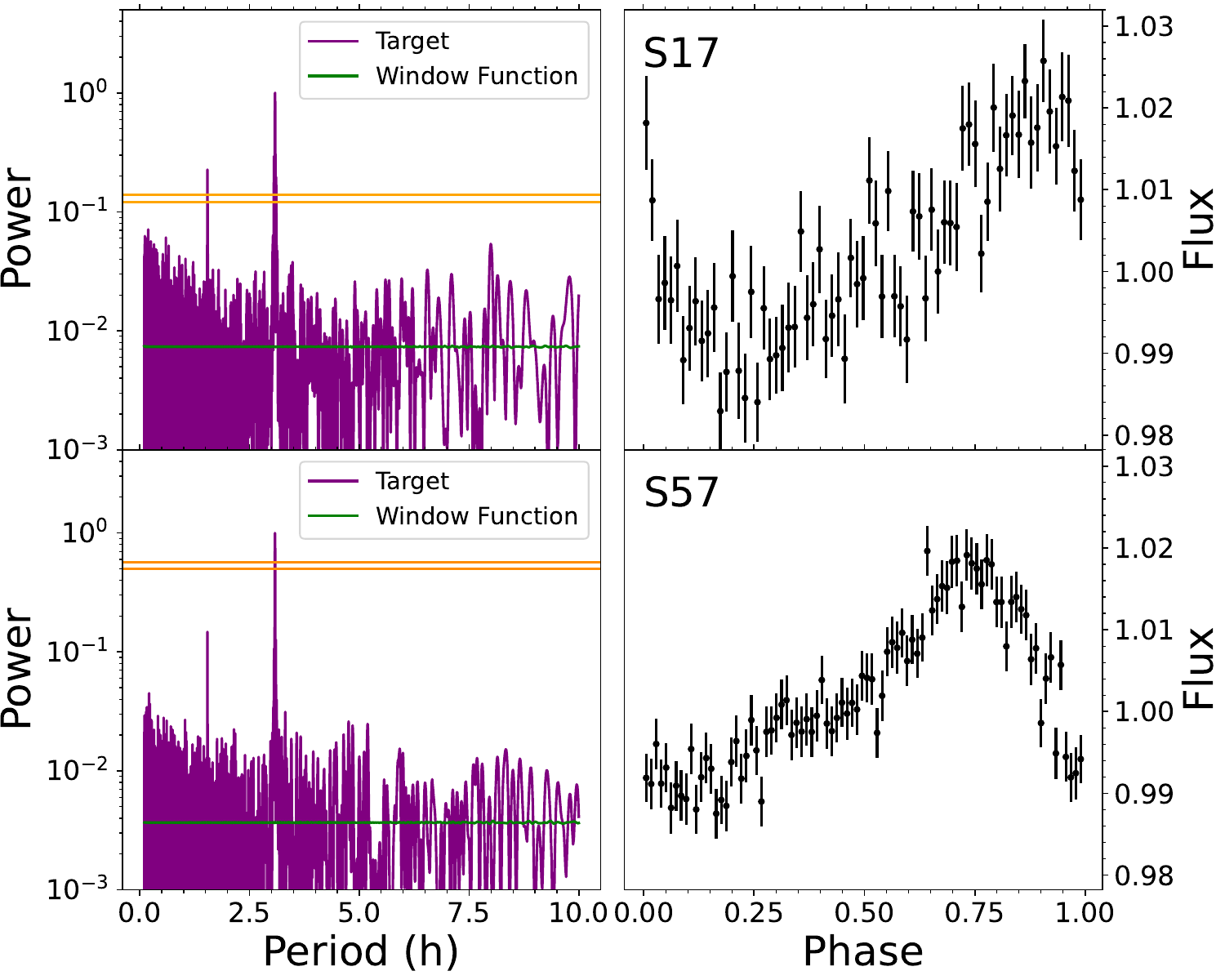}
            \caption{Lightcurves analysis for the 2 TESS sectors. Left panels show the LS periodogram for the S17 (top) and S57 (bottom). Horizontal orange lines show the 1\%  and 0.1\%  False Alarm Probability levels. Right panels show the lightcurves for sectors S17 (top) and S57 (bottom). We use 1500 points binning in the S17 data and 150 points binning in S57 as each dataset has a different length.}
            \label{fig:TESS}
    \end{figure} 
    
We used the GP kernel implemented in the {\sc celerite2} package \citep{celerite1, celerite2} and fit it to our data using a Markov Chain Monte Carlo using {\sc emcee} package \citep{2010CAMCS...5...65G, 2013PASP..125..306F}. This procedure was applied independently to sectors S17 and S57. The detrended TESS lightcurves phase-folded to the best-fit rotation period are shown in Figure ~\ref{fig:TESS}. The same periodicity is observed in both epochs with small morphology differences. We therefore derive a final rotation period from TESS data of P = $3.07908 \pm 0.00038$~h, calculated as the average of the best-fit values obtained for both sectors ($3.07870 \pm 0.00052$~h and $3.07946\pm 0.00055$~h for sectors S17 and S57, respectively). This period is also found by the Lomb Scargle periodogram shown in Figure ~\ref{fig:TESS}, where the second peak corresponds to the harmonic ($\sim$1.54~h). Both radio and optical periods are consistent within 1$\sigma$ and with previous measurements \citep{2008ApJ...684..644H, 2015ApJ...799..154M, 2026A&A...710A..22O}. The excellent agreement between the radio and optical periods indicates that both radio and optical variability are governed by the same rotation period. Similar agreement has previously been reported for the late-M dwarf TVLM 513-46546 \citep{2014ApJ...788...23W}, suggesting that his behavior may extend across the M/L transition.
    
\subsection{Radio lightcurves}
\label{subsec:Multi_LC}

\begin{figure}[h!]
        \centering
        \includegraphics[width=0.97\columnwidth]{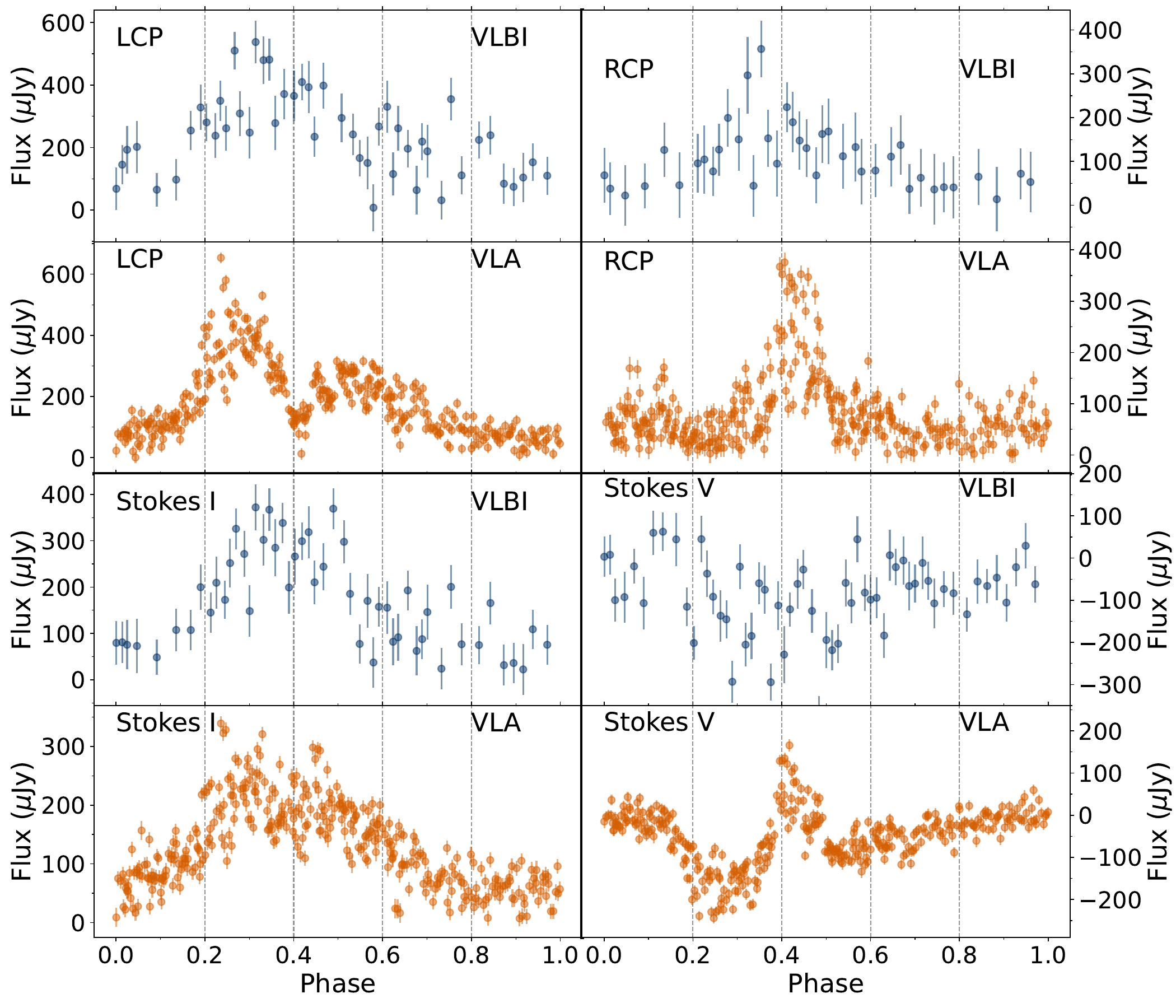}
        \caption{Radio lightcurves of LSPM J0036+1821. The data are separated into observations obtained with VLBI observatories (EVN and VLBA; top and third rows) and with the VLA (second and bottom rows). Left panels show the LCP (first and second rows) and Stokes I (third and fourth rows) parameter while right panels show RCP (first and second rows) and Stokes V (third and fourth rows) parameter. For the VLBI observations, all epochs were phase-folded using a rotation period P = 3.07941~h and averaged every 8 points. The VLA lightcurves are phase-folded using the same rotation period. Despite the observations being separated by several months to years, and obtained at different frequencies regimes, the main morphological features are consistent across datasets and occur at the same rotational phase. Vertical dashed gray lines represent 0.2 phase intervals.}
        \label{fig:radio_lc}%
    \end{figure}

The radio lightcurves are presented in Figure ~\ref{fig:radio_lc}. The high sensitivity of the VLA, which can achieve an rms of $\sim$15~$\mu$Jy for an integration time of 60~s in the whole 8 -- 12 GHz bandwidth, enables a very detailed characterization of the morphology of the pulses. To allow a more comprehensive analysis of the pulse morphology, we phase-folded the VLA lightcurves using the radio rotation period derived in Section~\ref{subsec:Periodic_Rad}. The LCP lightcurves exhibit a characteristic double-peak morphology, with two maxima separated by $\sim$0.25 in rotational phase; being the first peak consistently stronger than the second (see Figure ~\ref{fig:VLA_lc_ev}). Hereafter, we refer to the first and stronger peak, ocurring between phases 0.2 -- 0.4, as L1, and to the second and weaker peak, occurring between phases 0.5 -- 0.7 as L2 (see Fig. ~\ref{fig:radio_lc}). In contrast, the RCP emission displays a single dominant peak ocurring between the two LCP maxima in phase. The VLA data further reveal significant variability from one rotation to the next (see Fig. ~\ref{fig:VLA_lc_ev}). The L1 peak decreases in intensity from $\sim$600~$\mu$Jy during the first rotation to $\sim$300~$\mu$Jy in the last, whereas the RCP peak increases from $\sim$150~$\mu$Jy to 350~$\mu$Jy. In contrast, the L2 peak remains approximately constant at $\sim$240~$\mu$Jy over the three rotations (Figure ~\ref{fig:VLA_lc_ev}). The physical origin of this short timescales variability is discussed in Section   ~\ref{subsec:Auroral_Model}.

\begin{figure*}[h!]
        \centering
        \includegraphics[width=8.25cm]{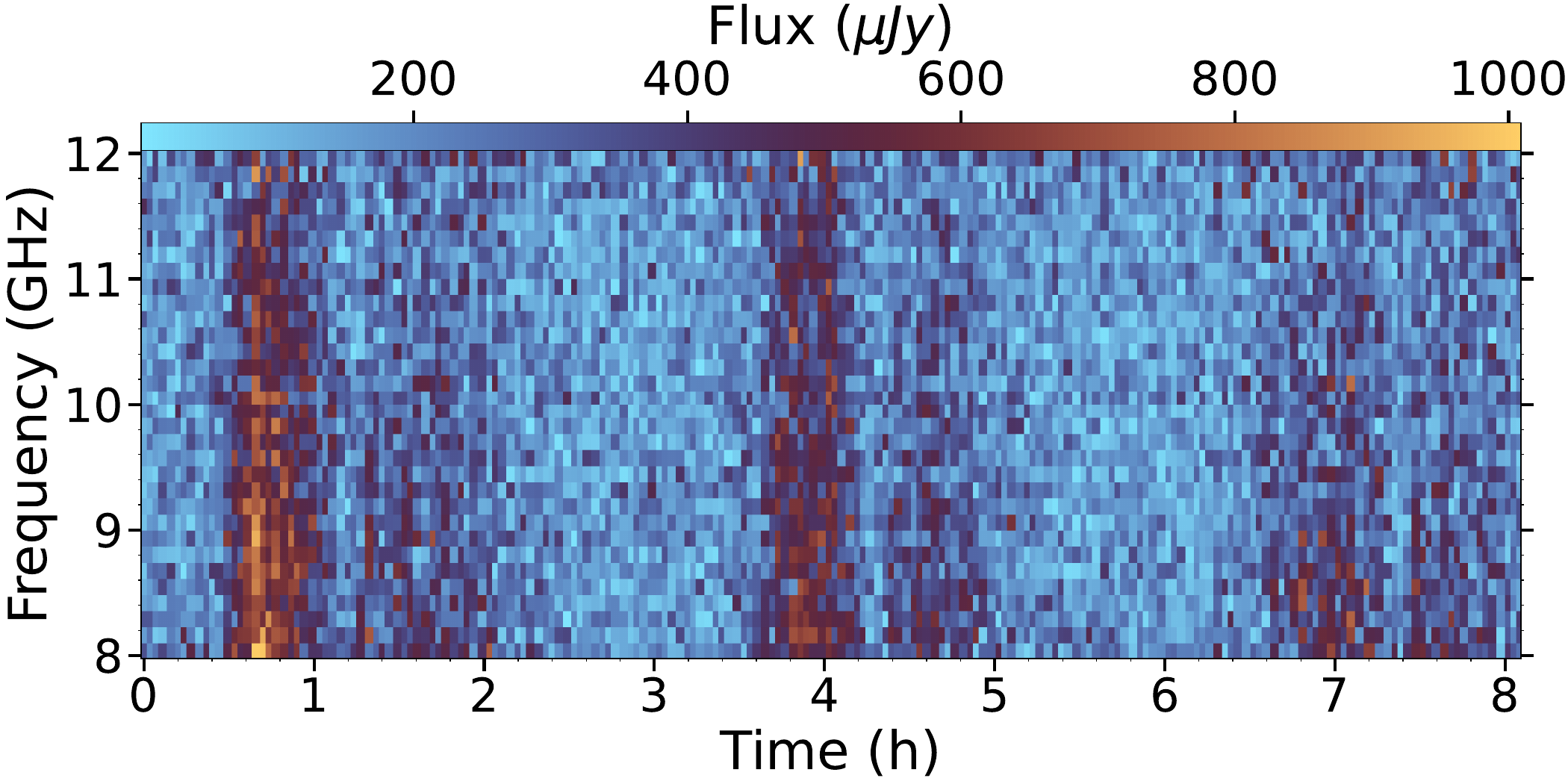}
        \includegraphics[width=8.25cm]{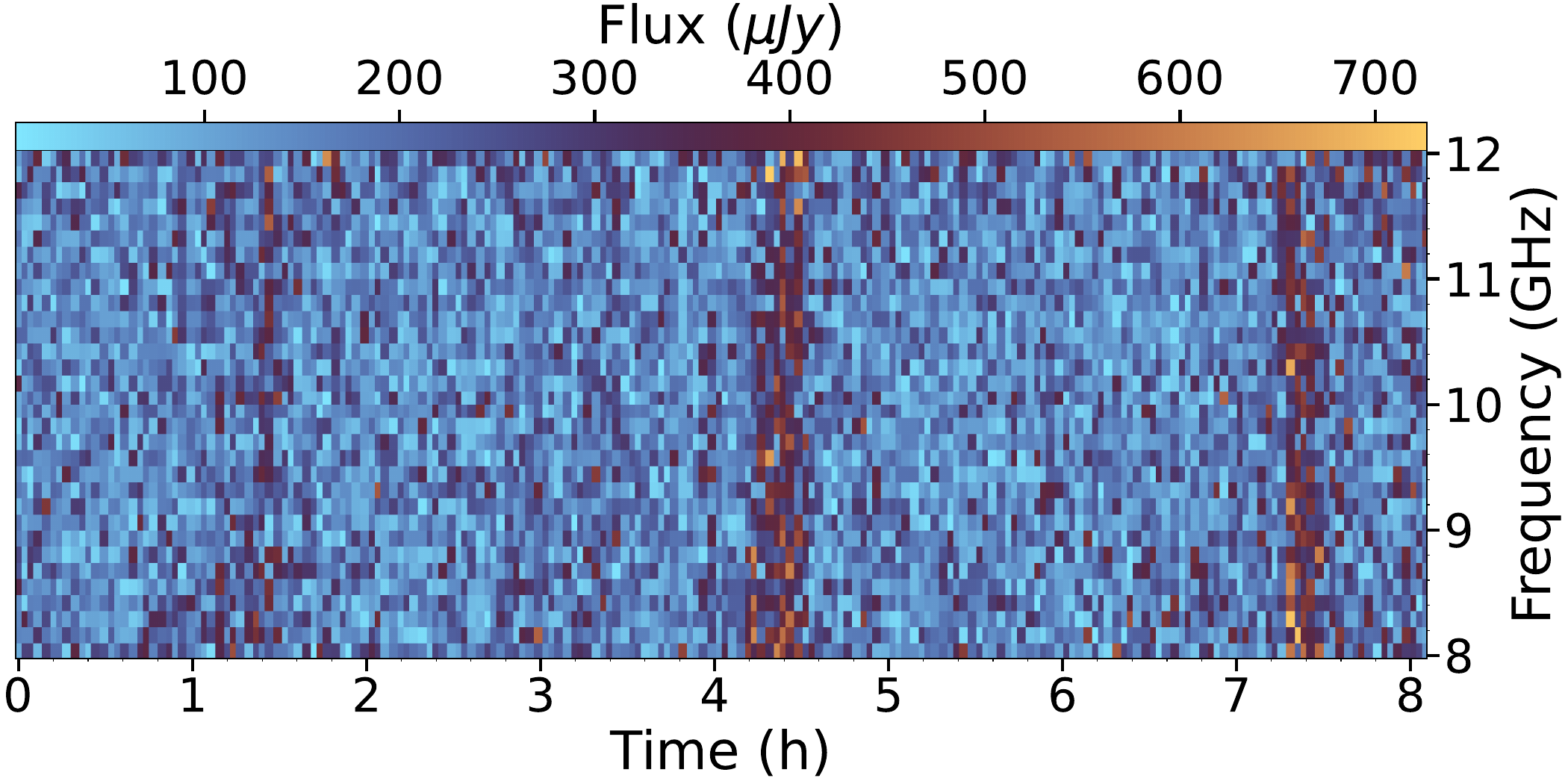}
            \caption{Dynamic Spectra of LCP (left panel) and RCP (right panel) VLA data in 8 - 12~GHz. We average the data across all baselines and used 128 MHz frequency bins and 60~s time binning. Pulses are clear at 0.8, 4 and 7 h.}
        \label{fig:dyn_spec}%
    \end{figure*}

Regarding the VLBI lightcurves (EVN and VLBA), we phase-folded the VLBI data using the previously derived rotation period of $3.07941$~h, adopting the start time of the earliest observing epoch (2019.17) as the reference phase. Although pulses are detected in each individual VLBI dataset, the relatively low signal-to-noise ratio (S/N) prevents a detailed characterization of their morphology (peak fluxes reach $\sim$200 - 500~$\mu$Jy meanwhile a typical VLBI lightcurve can deliver a mean rms of $\sim$150~$\mu$Jy for a 120~s integration time over the whole bandwidth, 128 MHz for the EVN and 512 MHz for the VLBA dataset). Instead, we averaged the data from all observing epoch, using a bin size of 8 data points ($\sim$4~min). With this procedure, we intend to better constrain the average pulse structure; as seen in Figure ~\ref{fig:radio_lc}, despite the individual VLBI datasets being acquired at different epochs, the corresponding averaged polarized lightcurves all reveal the same underlying structure. Remarkably, both VLBI and VLA lightcurves display a similar overall morphology. The agreement is less apparent in the Stokes V for two reasons. First, the Stokes V measurements have a lower S/N than the LCP and RCP data, a $\sqrt{2}$ factor. Second, because the VLBI observations span several years, small phase shifts of the dominant LCP pulse between epochs broaden its averaged profile. This broadening partially suppresses the contribution of the intermediate RCP pulse in the averaged Stokes $V$ profile, making it appear more different from the single-epoch VLA observations.

\subsection{Dynamic Spectra}
\label{subsec:Dyn_Spec}

We present in Figure \ref{fig:dyn_spec} the LCP and RCP dynamic spectra derived from the VLA observations. These spectra clearly trace the same emission features identified in the lightcurves shown in Figure ~\ref{fig:VLA_lc_ev}, a double-peaked structure in LCP and a single dominant peak in RCP. The strongest LCP peak (L1) exhibits hints of temporal evolution across successive rotations. Although the limited S/N prevents a detailed characterization, we tentatively observe changes in the emitting frequencies: while the overall burst intensity decreases from one rotation to the next, this variation appears to be more pronounced at higher frequencies, particularly toward $\sim$12~GHz. In contrast, the RCP dynamic spectrum shows a general increase in the intensity from one rotation to the next, with no clear evidence for a significant frequency dependence.

\section{Discussion}
\label{sec:Discus} 

\subsection{Radio emission of LSPM J0036+1821}
\label{subsec: Radio_em}
    
The pulsed emission shown in Figure ~\ref{fig:radio_lc} is clearly rotationally modulated and presents a high degree of circular polarization, which suggests a coherent mechanism. Both plasma emission and Electron Cyclotron Maser Instability (ECMI) are coherent mechanisms capable of producing high degrees of circular polarization. However, a fundamental distinction is that ECMI emission is highly beamed, being radiated in narrow hollow cones nearly perpendicular to the local magnetic field (Melrose \& Dulk 1982). When combined with the rotation, and potentially a non-zero magnetic obliquity, this strong beaming naturally produces bursty, periodically modulated lightcurves. The radio pulses presented in the Figure ~\ref{fig:radio_lc} are consistent with this scenario. This interpretation is well supported by previous studies, which have demonstrated that auroral radio emission in ultracool dwarfs is most naturally explained by ECMI \citep[e.g.,][]{2008ApJ...684..644H, 2018ApJS..237...25K}.

ECMI emission is a wave--particle interactions involving anisotropic electron distributions along magnetic field lines connected to high magnetic latitudes. Electrons with small pitch angles precipitate into the atmosphere, while those with larger pitch angles can be magnetically reflected or accelerated upward, producing the anisotropy required to drive maser amplification near the local electron cyclotron frequency. \citep{2023JGRA..12831985L, https://doi.org/10.1029/2024GL114447}. The emission occurs near the fundamental or second harmonic of the local cyclotron frequency, which directly relates the observed frequency to the magnetic field strength following the equation $f (GHz) = 2.8 \cdot B $ (kG). The dynamic spectra presented in Figure ~\ref{fig:dyn_spec} does not show a clear frequency cutoff in the observed frequencies (8 -- 12~GHz), despite such a cutoff being a characteristic feature of ECMI emission. This suggests that the cutoff frequency lies above the upper limit of the observed band; therefore we adopt 12~GHz as a lower limit to the maximum emission frequency, and we infer a corresponding lower limit on the magnetic field strength of $B > 4.3$~kG.

   \begin{figure*}[h!]
        \centering
        \includegraphics[width=14cm]{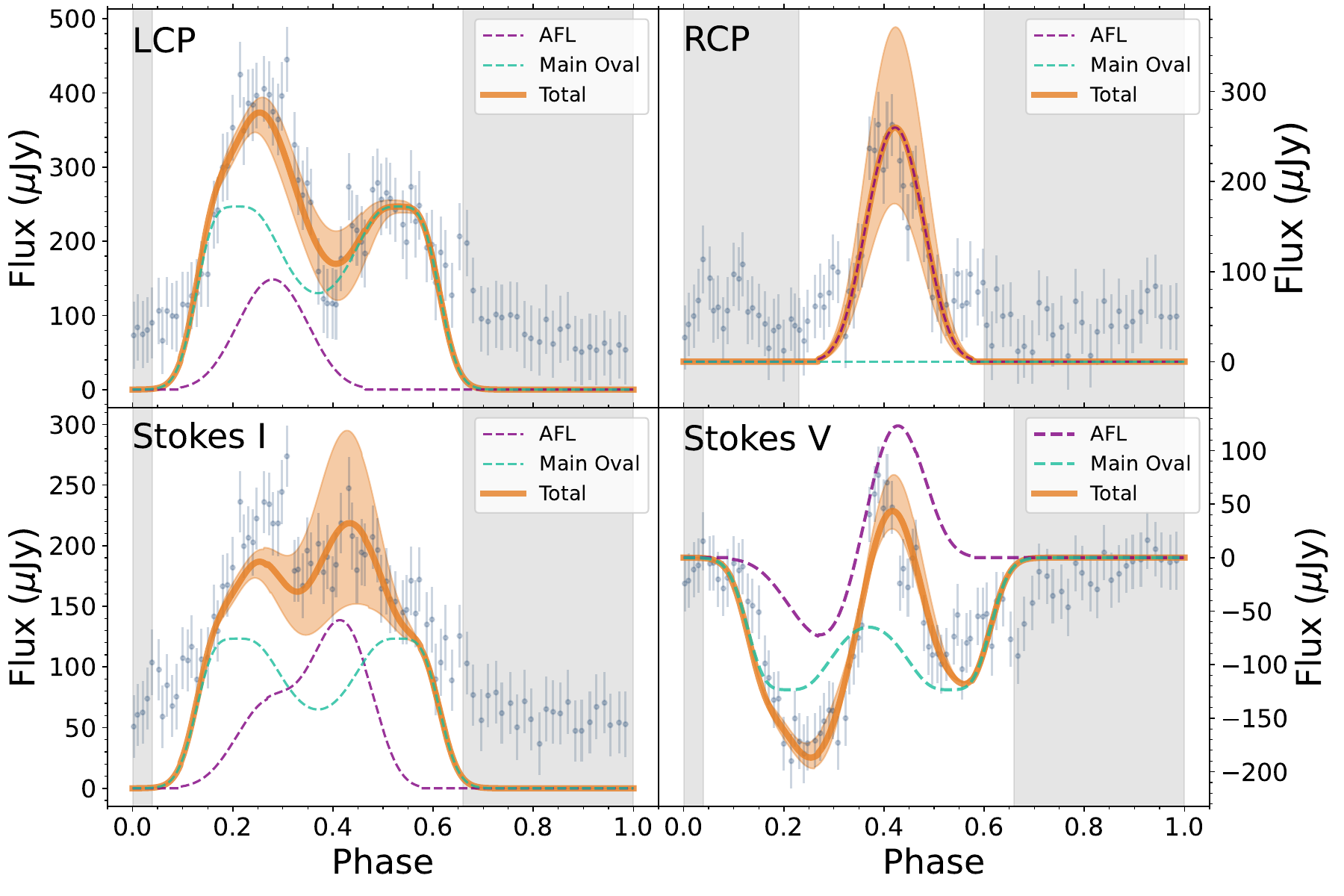}
        \caption{Model fit to the binned phase-folded VLA radio lightcurves of LSPM J0036+1821. The top panels show the circularly polarized components, LCP (left) and RCP (right), while the bottom panels display the corresponding Stokes I (left) and Stokes V (right) lightcurves. The best-fit model, corresponding to the parameters listed in Table~\ref{table:AURO_MODEL}, is overplotted. The VLA binned data are shown in dark blue and the 1$\sigma$ error bars are plotted. The contribution from the active field line component is shown as a purple dashed line, the main oval component is shown in cyan dashed line. The complete model, obtained as the sum of both components, is plotted in an orange solid line. The 3$\sigma$ variance of the model is plotted as a shadowed orange region. Gray shaded regions indicate phases dominated by non-auroral emission, which are not included in the model fitting.}
        \label{fig:radio_fit}%
    \end{figure*}

\subsection{A Jupiter-like auroral model}
\label{subsec:Auroral_Model}

Auroral emission from Jupiter has been extensively studied across multiple wavelength regimes, including the ultraviolet \citep[e.g.,][]{1996Sci...274..404C}, infrared \citep[e.g.,][]{2025NatCo..16.3907N}, and radio \citep[e.g.,][]{2017A&A...604A..17M}. The auroral radio emission is driven by two principal mechanisms: (i) high-latitude auroral emission encircling the magnetic poles (main-oval emission), and (ii) electrodynamic interactions between Jupiter and its moons, most prominently with Io.

The main oval is associated with the presence of a co-rotating plasma disk within the magnetosphere of Jupiter. Beyond a certain distance \citep[$\sim$20~$\rm R_{Jup}$;][]{1979GMS....21..297H}, the breakdown of rotational coupling between the magnetosphere and the ionosphere generates currents which flow along field lines from the plasma sheet to high ionospheric latitudes and back to it. These field-aligned currents accelerate electrons toward the upper atmosphere, providing the necessary component to feed the ECMI auroral emission \citep[see][for details]{2001P&SS...49.1067C}.

The second major auroral mechanism in Jupiter arises from its interaction with its orbiting moons \citep[e.g.,][]{1964Natur.203.1008B, 2017JGRA..122.9228L, 2017pre8.conf...45Z, 2007JGRA..112.5213H}. This process is driven by the motion  of the satellites relative to the Jupiter's magnetized plasma torus. In this situation, the moon generates electric currents and Alfv\'en waves along the magnetic field lines that connect the satellite to the planet. In a manner analogous to the main oval mechanism, these field-alligned currents accelerate electrons toward Jupiter's high-latitude atmosphere, where, at radio wavelengths, intense emission is generated via the ECMI mechanism \citep{2023JGRA..12831985L}. The radio emission is generated along magnetic field lines associated with the main Alfv\'en wing, which are commonly referred to as active field lines (AFL).

The presence of two distinct and asymmetric pulses in the LCP lightcurves is not successfully reproduced by either the main oval or the active field line component alone, as each mechanism individually tends to generate symmetric lightcurves with respect to a certain rotation phase \citep[e.g., ][]{2017ApJ...846...75P}. On the other hand, an arbitrary number of active field lines could reproduce the observed lightcurves but at the cost of overparameterizing the model. Assuming the similarity with the Jupiter-Io system, we therefore adopt a hybrid approach combining the emissions from both the main auroral oval and one active field line. Specifically, we implement the geometrical models introduced by \cite{2024A&A...682A.170B} for the main-oval case and by \cite{2024A&A...692A..66K} for the case of the active field line. A detailed description of these model geometry is provided in the previous references, here we summarize their key assumptions (see also Appendix \ref{sec:Ap:model}). ECMI emission is assumed to be radiated within a hollow cone characterized by an opening angle $\alpha$ and a thickness $\Delta \alpha$,  and it is detectable when the observer's line of sight lies within the angular range $\alpha \pm \Delta \alpha/2$. In the main oval model, emission originates from an auroral ring encircling the magnetic poles, where each location along the ring acts as an ECMI emitter with an identical hollow cone beaming geometry. In the active field line model, only one or a limited number of magnetic field lines act as localized ECMI emitters, analogous to the Jupiter-Io interaction. In both models, emission from both hemispheres is assumed to arise at symmetric magnetic latitudes and to share the same beaming pattern, differing only in circular polarization handedness.

The models depend on a set of geometrical parameters. The inclination of the rotation axis relative to the observer's line of sight (i), the magnetic obliquity ($\beta$) and the rotational phase offset ($\phi_{\rm 0}$) are intrinsic properties of the system and therefore common to both model components. For the main oval, additional parameters include the magnetic co-latitude of the ring $\theta_{\rm MO}$, the cone opening angle $\alpha_{\rm MO}$, its thickness $\Delta \alpha_{\rm MO}$, and flux scaling factor $\rm F_{\rm MO}$. The active field line component adopts a similar parametrization, where $\theta_{\rm AFL}$ describes the magnetic co-latitude of the ECMI emitter, $\varphi$ represents the magnetic longitude of the active field line, and $\alpha_{\rm AFL}$, $\Delta \alpha_{\rm AFL}$ and $\rm F_{\rm AFL}$ define the beaming geometry and flux scaling, similarly to the main oval component. Finally, the parameter $\rho$ defines the gaussian profile for both main oval and active field line emission (see Appendix $\ref{sec:Ap:model}$).

ECMI emission can be produced in either extraordinary (x-mode) or ordinary (o-mode) modes. For a given mode, the observed polarization depends on the direction of the local magnetic field relative to the observer. Under the IAU convention and the magnetic geometry adopted here, x-mode emission has positive/negative Stokes V in the northern/southern magnetic hemisphere, respectively, with the opposite behavior for o-mode. If the plasma density $v_{\rm p}$ is lower than the cyclotron frequency $v_{\rm B}$, such as $v_{\rm p}/v_{\rm B} < 0.05$ x-mode dominates \citep{1982AuJPh..35..447H} although the o-mode could still be present at a lower intensity. Indeed, theoretical expectations for ultracool dwarf magnetospheres favor x-mode emission with o-mode being partially or complete supressed \citep{2011A&A...526A.161K}. We therefore adopt x-mode as the preferred scenario.

\begin{table}[h!] 
\centering
\caption{Prior and posterior distributions of the auroral-model parameters.}
\label{table:AURO_MODEL}
\begin{tabular}{ccc}
\hline\hline             
Parameter & Prior & Posterior \\
\hline
  i (º) & Fixed (97) & Fixed (97)\\
  $\beta$ (º) & U(0, 60) & $3.6 \pm 0.2$\\
  $\alpha_{\rm AFL}$ (º) & U(70, 90) & $89.02 \pm 0.08$ \\
  $\alpha_{\rm MO}$ (º)& U(70, 90) & $85.01 \pm 0.15$\\
  $\Delta\alpha_{\rm AFL}$ (º)& U(1, 15) & $9.92 \pm 0.21$\\
  $\Delta\alpha_{\rm MO}$ (º)& U(1, 15) & $5.6 \pm 0.4$\\
  $\theta_{\rm AFL}$ (º)& U(0, 40) & $1.15 \pm 0.07$\\
  $\theta_{\rm MO}$ (º)& U(0, 40) & $0.9 \pm 0.06$\\
  $\phi$ (º)& U(0, 360)& $304.4 \pm 2.7$\\
  $\varphi_{\rm 0}$& U(0, 1) & $0.630 \pm 0.001$\\
  $\rm F_{AFL}$ (mJy)& U(0.1, 10) & $2.1 \pm 0.5$\\
  $\rm F_{MO}$ (mJy)& U(0.1, 10) & $0.463 \pm 0.014$\\
\hline
\end{tabular}
\tablefoot{The AFL and MO subscripts refer to the Active Field Line and Main Oval components, respectively.}
\end{table}
\subsubsection{Implementation of the model}

There is currently limited observational information available to constrain the geometrical parameters specified above. Measurements of $v$\,sin\,$i$ for LSPM J0036+1821 have been reported by \cite{2010ApJ...723..684B}, \cite{2008ApJ...684.1390R}, \cite{2006ApJ...647.1405Z}, \cite{2005MNRAS.358..105J} and \cite{2017ApJ...842...78V}. We adopt an average $v$\,sin\,$i$ value of $36.4 \pm 1.2$ km/s. Following the method presented in \cite{2020AJ....159...81M}, assuming a radius of $1.01 \pm 0.07$ $\rm R_{\rm Jup}$ \citep{2015ApJ...810..158F} and using the derived radio rotation period, we derive an inclination of i = $65 \pm 15$ deg. Since spectroscopic measurements cannot distinguish between an inclination i and its supplementary angle, both configurations yield the same projected velocity. Accordingly, we must therefore consider in our 3D model not only the inclination derived directly from $v$\,sin\,$i$, but also its supplementary angle. 
We first used the nested sampling Monte Carlo algorithm MLFriends (Buchner, 2014; 2019) implemented in the UltraNest\footnote{\url{https://johannesbuchner.github.io/UltraNest/}} package (Buchner 2021), to explore a wide range of the parameter space. In a second step, the a priori UltraNest solution was used to initialize a more flexible and faster Markov Chain Monte Carlo fit performed with the {\sc emcee} package \citep{2010CAMCS...5...65G, 2013PASP..125..306F}. This strategy allows a deeper and more accurate determination of the model parameters. A summary of the adopted prior can be found in Table ~\ref{table:AURO_MODEL}.

   \begin{figure*}[h!]
        \centering
        \includegraphics[width=7.cm]{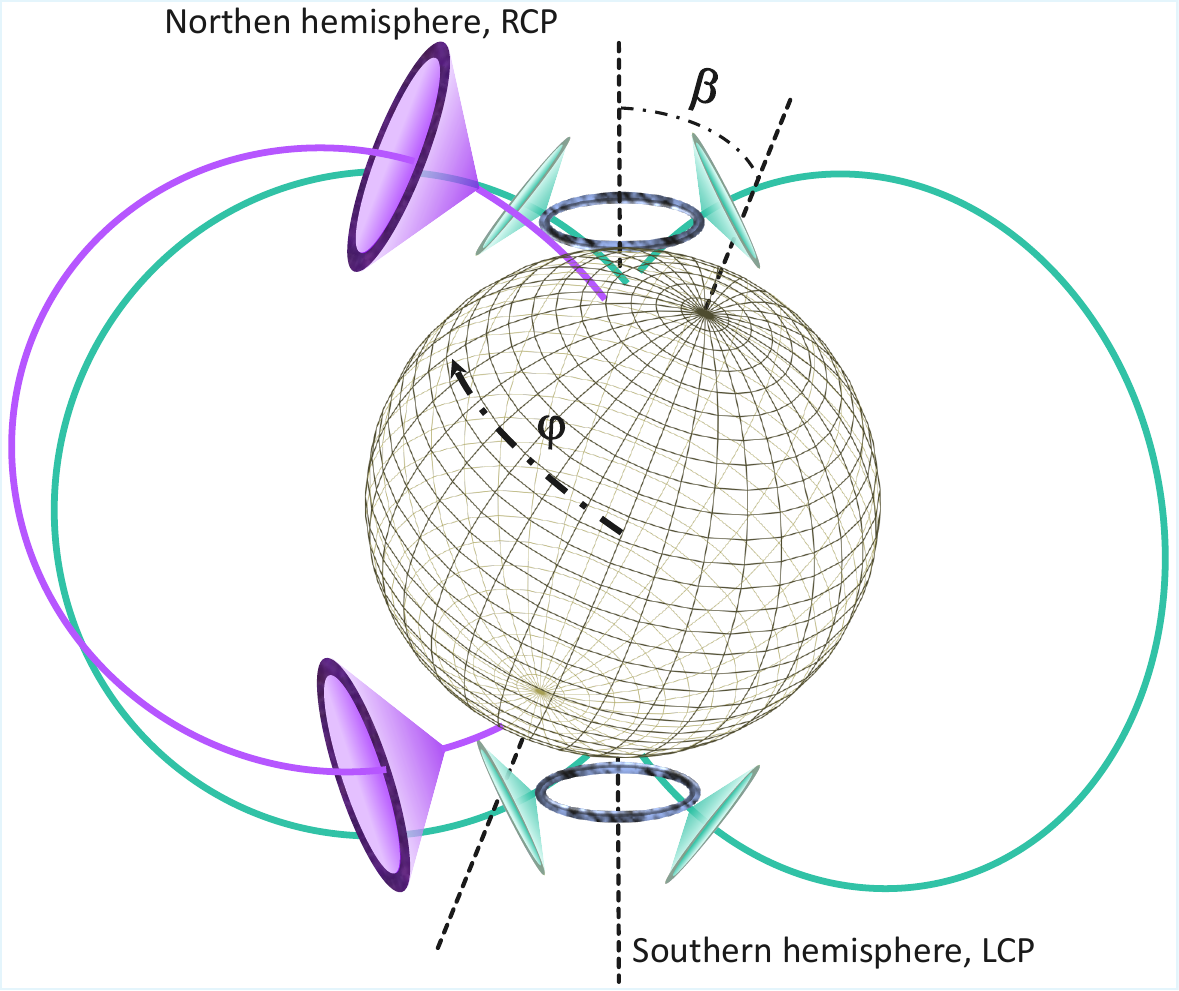}
        \includegraphics[width=7.cm]{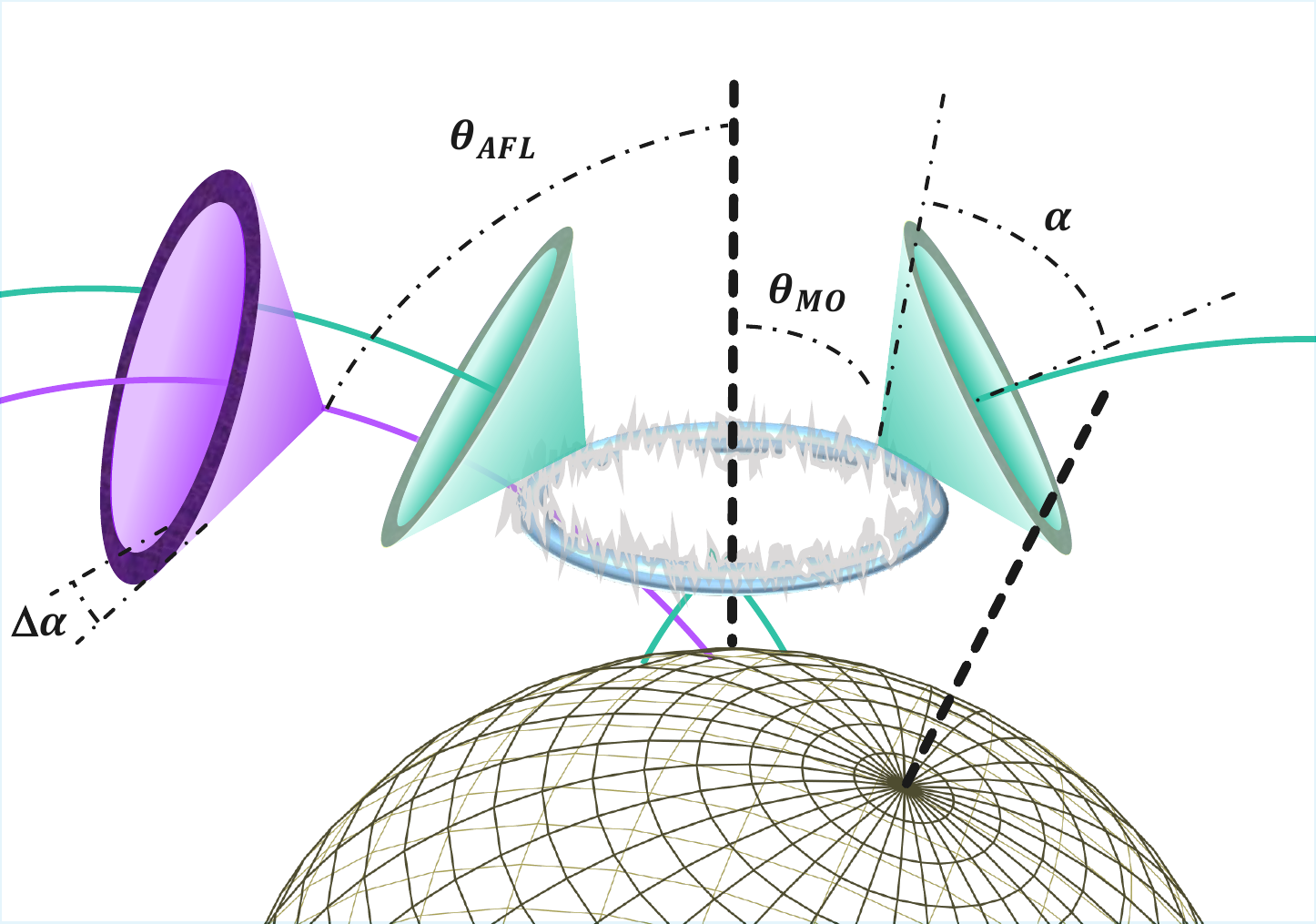}
        \caption{Geometric configuration of LSPM J0036+1821 using the parameters derived from the model fit, shown in Table ~\ref{table:AURO_MODEL}. The left panel illustrates the large-scale magnetospheric geometry. The structures responsible for the emission are shown in different colors: the active field line component is plotted in purple while the main oval component is shown in cyan. In the case of the main oval, each point along the auroral ring acts as an ECMI emitter; however, only two representative emission cones are shown for clarity. Note that both magnetic poles have equivalent geometries, differing only in the handedness of the emission. The right panel presents a zoomed view of one magnetic pole. The illustrations are not to scale, as it is intended for schematic purposes only.}
        \label{fig:Geo_conf}%
    \end{figure*}

We note that the model exhibits intrinsic degeneracies, particularly between (i) the inclination angle $i$ and the magnetic obliquity $\beta$, and (ii) the parameter $\rho$ and the cone thickness $\Delta\alpha$. The initial UltraNest exploratory run yields cos i = $-0.12 \pm 0.04$, corresponding to an inclination angle of $97^\circ \pm 3^\circ$, consistent with the observational constrains obtained from the $v$\,sin\,$i$ measurements. With respect to $\rho$, the a priori value provided by UltraNest is $\rho = 7.3 \pm 0.3$, which implies that at the boundary of the wall cone the flux is 1000 times lower than the peak flux, as expected by the ECMI theory. Consequently, we fix both parameter in the MCMC fitting to reduce the impact of the geometrical degeneracies mentioned.

\subsubsection{Estimate of model parameters}

The posterior distribution of the parameters after our MCMC analysis for the phase-folded VLA data is presented in Table ~\ref{table:AURO_MODEL}, the corner plot is shown in Figure ~\ref{fig:CP_MEANMODEL}. The best-fit solution is shown in Figure ~\ref{fig:radio_fit} and the corresponding geometry is illustrated schematically in Figure ~\ref{fig:Geo_conf}. The fit favors an almost equator-on viewing configuration, using the supplementary inclination allowed by the $v$\,sin\,$i$ constraint. The inferred magnetic obliquity is small, $\beta = 3.6 \pm 0.2$º, so the large-scale dipole is close to aligned with the rotation axis. Even a small obliquity is sufficient to modulate the visibility of narrow ECMI beams because the emission is observed only when the line of sight intersects the cone wall. Both emitting components are located at high magnetic latitude. The active field line solution gives $1.15 \pm 0.07$º, while the main oval gives $0.87 \pm 0.06$º. These values should be interpreted as effective, band-averaged source co-latitudes within a dipolar geometric model, rather than as a direct measurement of a unique physical altitude. For the beam geometry we derive a half opening angle of $89.16 \pm 0.08$º for the active field line component and $85.01 \pm 0.15$º for the main oval, as expected for ECMI emission ($\alpha \sim70 - 90 $º). We also derive a cone thickness of $9.9 \pm 0.2$º and $5.6 \pm 0.4 $º for the active field line and the main oval, respectively. Because of the finite-window prescription adopted here, these correspond to emission extending approximately $\sim$5$^\circ$ and $\sim$2.8$^\circ$ around the preferred emission direction ($\alpha$) for the active field line and the main oval, respectively. 
These thicknesses may appear larger than the $\sim$1--2$^\circ$ cone-wall widths commonly inferred for Jovian decametric emission \citep{2016A&A...596A..18P}. However, ECMI emission is in fact composed of thousands of millisecond-duration bursts \citep[e.g.,][]{2023NatCo..14.5981M}, which collectively appear at macroscopic scales as emission arcs. Therefore, the observed width of an emission arc reflects the longitudinal extent of the emitting region. For Jupiter and Saturn, this longitudinal extent is typically $\sim$1--2$^\circ$. In our case, the VLA light curves fitted here are averaged over the full 8--12~GHz band. Since the ECMI frequency is tied to the local cyclotron frequency, the observed band samples a finite range of magnetic-field strengths and therefore a finite range of source positions along the emitting field line. In addition, the half-opening angle of the ECMI cone is expected to vary with frequency \citep{2008GeoRL..3513107H}. The fitted $\Delta\alpha$ should therefore be regarded as an effective angular width that absorbs unresolved frequency-dependent beaming, finite source extent, and any residual intrinsic spread of the cone wall.

Figure ~\ref{fig:radio_fit} shows the model lightcurves obtained using the parameters listed in Table ~\ref{table:AURO_MODEL}. In this framework, the main oval provides the broad, relatively stable LCP contribution, including most of the L2 peak and part of L1. For the best-fit geometry its visible contribution is dominated by the southern magnetic hemisphere, and therefore it appears mainly as LCP under the x-mode convention. The AFL accounts for the stronger and more variable L1 feature and for the RCP peak.

\subsection{Short-term evolution of the lightcurves}
\label{subsec:LC_tempevol}

The VLA lightcurves reveal a clear temporal evolution between successive rotations (see Figure ~\ref{fig:VLA_lc_ev}). As described in Section ~\ref{subsec:Multi_LC}, the RCP and LCP peaks attributed to the active field line exhibit an anticorrelated amplitude evolution, with one component strengthening as the other weakens. To disentangle the origin of this behavior, we attempt to model the temporal variability by identifying which parameters of the active field line framework could plausibly change on these time scales. We assume that the main oval contribution remains constant as the second LCP peak (L2) is relatively stable over the observations. Large-scale geometrical parameters such as the inclination angle and magnetic obliquity are not expected to vary over the $\sim$9~h duration of the VLA observations. In contrast, parameters associated with the location or beaming of the radio emission are more likely to change. 

   \begin{figure*}[h!]
        \centering
        \includegraphics[width=15cm]{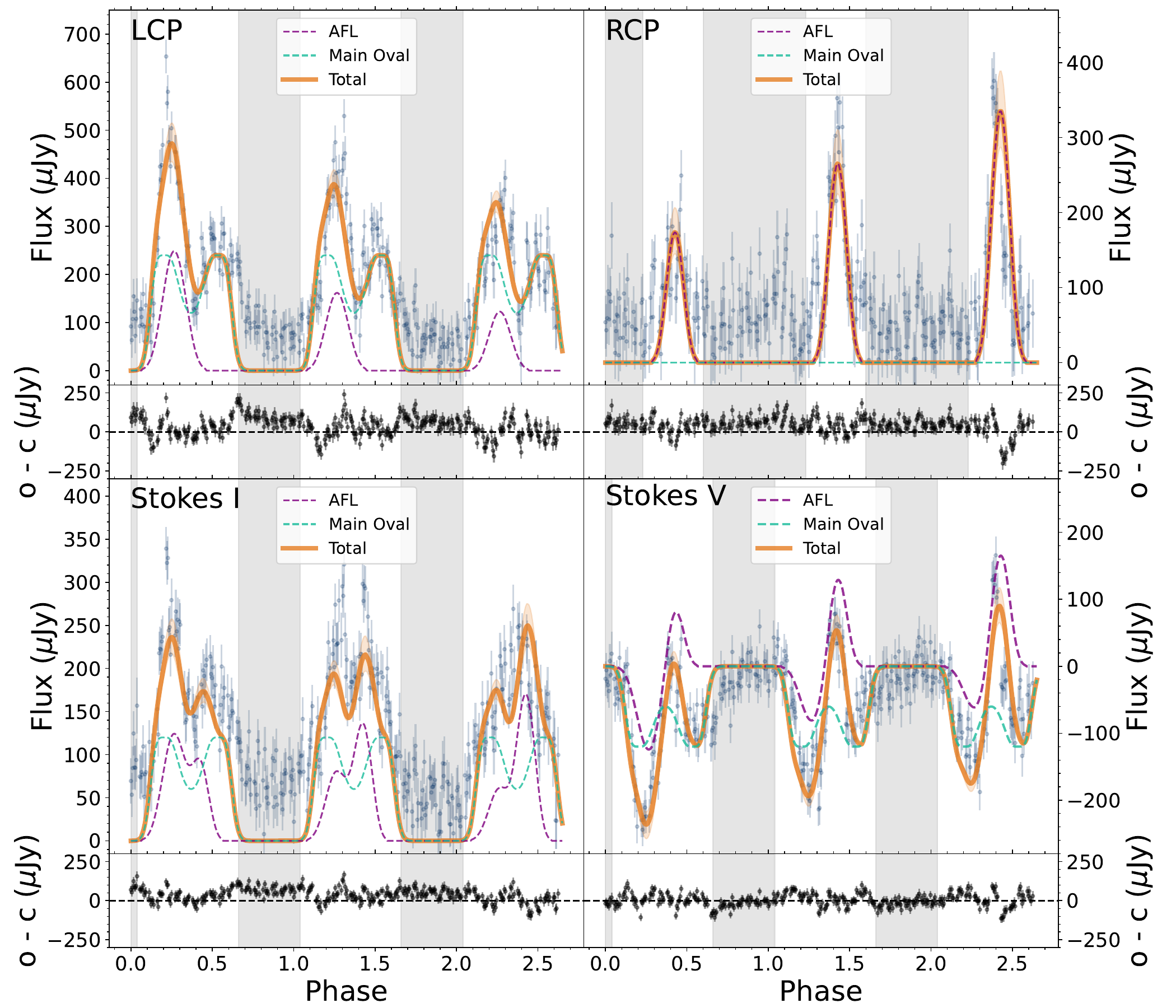}
        \caption{Model fitting for the VLA radio lightcurves. In the top panel, LCP (left) and RCP (right) light curves are shown, while the bottom panel displays Stokes I (left) and Stokes V (right). Data are shown as blue dots. The best-fit model corresponding to the parameters listed in Table~\ref{table:AURO_MODEL} for all the parameters except for $\alpha$ is overplotted. For this parameter we change its value for each rotations according to the values shown in the Table ~\ref{table:AURO_MODEL_ROT}. The contribution from the Active Field Line (active field line) component is shown in purple, the Main Oval component in cyan, and the sum of both components in orange. Gray shadowed regions correspond to non-auroral emission, the model is not suppose to model this parts. Black data points shown below each panel represent the residuals of the fit (data - model).}
        \label{fig:radio_fit_ev}%
    \end{figure*}

We explored how variations in different parameters affect the modeled lightcurves; namely: (i) changes in the co-latitude primarily produce a phase shift in the LCP ligthcurve, along with small intensity variations in the same direction for both LCP and RCP; (ii) variations in the half-opening angle $\alpha$ lead to an anticorrelated evolution of the LCP and RCP peak amplitudes; (iii) changes in the cone thickness $\Delta \alpha$ result in correlated growth/decrease of LCP and RCP components; and (iv) the longitudinal position of the active field line $\varphi$ produce structural changes in the lightcurves similar to those caused by variations in the opening angle. Among the parameters explored, those that best reproduce the observed rotation-to-rotation variability are the averaged half-opening angle $\alpha$ and the longitudinal position $\phi$, as both can produce the anticorrelated evolution between LCP and RCP peaks. Given this degeneracy, we decide to explore changes in both parameters independently.

Additionally, if the symmetry between the two magnetic hemispheres is relaxed, allowing each pole to vary independently, changes in the electron beam density could also produce an anticorrelated evolution of the LCP and RCP fluxes, reproducing the observed morphology. However, within the framework of our geometric model, the flux acts only as a scaling factor, making such a scenario largely ad-hoc. Exploring this possibility would therefore require a more sophisticated model that explicitly incorporates the physical processes encapsulated by the flux parameter, which is beyond the scope of this work.

\subsubsection{Active Field Line's half-opening angle variability}
\label{subsubsec:alpha_var}

We explored the variability driven by changes in the averaged half-opening angle $\alpha$. We fit data using {\sc emcee} MCMC sampler, fixing all parameters with the values listed in Table ~\ref{table:AURO_MODEL} and allowing $\alpha$ to vary independently for each rotation. A summary of the adopted priors and resulting posterior distributions distribution is presented in Table ~\ref{table:AURO_MODEL_ROT}.

\begin{table}[h!]
\centering
\caption{Prior and posterior distributions of the rotation-dependent auroral-model parameters.}
\label{table:AURO_MODEL_ROT}

\begin{tabular}{ccc}
\hline\hline
Parameter & Prior & Posterior \\
\hline
$\alpha_{\rm AFL,1R}$ (º) & U(70, 90) & $88.8 \pm 0.01$\\
$\alpha_{\rm AFL,2R}$ (º) & U(70, 90) & $89.05 \pm 0.01$\\
$\alpha_{\rm AFL,3R}$ (º) & U(70, 90) & $89.20 \pm 0.01$\\
\hline
\end{tabular}
\tablefoot{The model allows $\alpha_{\rm AFL}$ to vary between consecutive rotation periods.}
\end{table}

We derive a cone half-opening angle of $88.8 \pm 0.01$º for the first rotation, followed by an increase of 0.2º for the second rotation and 0.15º for the third rotation. The resulting lightcurves with the parameters derived allowing $\alpha$ to change between rotations are shown in Figure ~\ref{fig:radio_fit_ev}. With this approach, we successfully reproduce the observed anticorrelated evolution of the LCP and RCP peaks by introducing only small changes in the emission cone opening angle from one rotation to the next.

The half-opening angle of the ECMI hollow cone depends on the emission frequency range, the maximum cyclotron frequency and the electrons velocity, as described by Equation \ref{ec:freq_dep} \citep{2008GeoRL..3513107H, 2026A&A...705A.149C}:

\begin{equation}
\label{ec:freq_dep}
\alpha = \arccos \left[\frac{\Gamma^{-1}}{N}\frac{\rm v/c}{(1 - \frac{f_{\rm ce}}{f_{\rm ce, max}})^{1/2}} \right]
\end{equation}

Being $\rm v$ the velocity of the electrons, $\Gamma$ the Lorentz factor, N the local refractive index, $f_{\rm ce}$ the local emission cyclotron frequency and $f_{\rm ce, max}$ the maximum cyclotron emitting frequency along the active field line.

The derived emission cone half-opening angle represents an average value across frequencies, as Equation ~\ref{ec:freq_dep} shows that the cone geometry is intrinsically frequency dependent. Although the dynamic spectra presented in Figure ~\ref{fig:dyn_spec} do not allow a detailed characterization of the frequency dependence of the emission, we observe hints of rotation-to-rotation changes in the spectral extent of the bursts. Such tentative changes in the frequency distribution could lead, according to Equation ~\ref{ec:freq_dep}, to corresponding variations in the average opening angle of the emission cone, thereby contributing to the observed temporal evolution of the pulse morphology. A similar feature is observed in the Jupiter-Io interaction, specifically in Jupiter's Io Decametric Radio Emission (Io-DAM), where the upper ECMI cutoff frequency depends on Io's orbital position \citep[see Fig. 13][]{2017A&A...604A..17M}. Variations in the ECMI cutoff frequency can consequently lead to changes in geometry of the emission cone. Indeed, following \cite{2019GeoRL..4611606L} and \cite{2026A&A...705A.149C}, variations in the cone opening angle may result from either (i) changes in the local magnetic-field configuration at the footprint of the active magnetic field line or (ii) variations in the local plasma density, possibly associated with a local magnetic reconfiguration or the injection of a plasmoid.

\subsubsection{Active Field Line's longitude variability}
\label{subsubsec:phi_var}

The longitude of the active field line could also produce an anticorrelated change in the LCP/RCP lightcurves. Following the same approach as that described for $\alpha$, we investigate how changes in $\phi$ can reproduce the morphology variability. For that purpose we fit data using {\sc emcee} MCMC sampler, fixing all parameters with the values listed in the Table ~\ref{table:AURO_MODEL} and allowing $\phi$ to vary independently for each rotation. The adopted priors and posterior distribution can be found in Table ~\ref{table:AURO_MODEL_ROT_PHI}. We derive a longitude position of the active field line of $297.3 \pm 0.2$º for the first rotation, followed by an increase of $\sim$7º and $\sim$8º for the second and third rotation, respectively. The corresponding fit is shown in Figure ~\ref{fig:model_phi} and reproduces the main features observed in the lightcurves morphology across the three rotations.
\begin{table}[h!] 
\centering
\caption {Prior and posterior distributions of the rotation-dependent auroral-model parameters.}
\label{table:AURO_MODEL_ROT_PHI}
\begin{tabular}{ccc}
\hline\hline             
Parameter & Prior & Posterior \\
\hline
  $\phi_{\rm AFL, 1R}$ (º) & U(0, 360) & $297.3 \pm 0.2$\\
  $\phi_{\rm AFL, 2R}$ (º)& U(0, 360) & $304.2 \pm 0.3$\\
  $\phi_{\rm AFL, 3R}$ (º)& U(0, 360) & $312.8 \pm 0.3$\\
\hline
\end{tabular}
\tablefoot{The model allows $\phi_{\rm AFL}$ to vary between consecutive rotation periods.}
\end{table}
Under the hypothesis that the active field line is caused by the presence of a planetary companion (on analogy with the Jupiter-Io system), changes in the longitude of the active field lines would naturally be explained by the orbital motion of a planet around LSPM J0036+1821. We explore this possibility in turn.

\subsubsection{Origin of the Active Field Line}
\label{subsec:active field line}

Our data are well reproduced by a model consisting of a main oval and an active field line (see Fig. ~\ref{fig:radio_fit}). The rotation-to-rotation variability observed in the lightcurves can be explained by small changes in either the half-opening angle of the emission cone or the longitude of the active field line (Figure ~\ref{fig:radio_fit_ev} and Figure ~\ref{fig:model_phi}). 
\cite{2008ApJ...684..644H} reported 3.5 consecutive rotational cycles of this same target and identified a very similar overall morphology. In their data, the brightest LCP and RCP peaks also exhibited an anticorrelated amplitude evolution, with one component strengthening as the other weakened \citep[see Fig. 2][]{2008ApJ...684..644H}. The stability of the lightcurves, despite the $\sim$15 years separation between the observations, suggests that the overall emission geometry may remain stable on timescales of years. The persistence of this characteristic morphology over the years points toward a recurring or quasi-periodic mechanism modulating the auroral emission. 

One possible explanation for the observed periodic modulation is the presence of a companion in the sub-Alfv\'enic region (around a few stellar radii), which could activate the magnetic field line and modulate the auroral emission with its orbital period, analogous to the Jupiter-Io electrodynamic interaction \citep{2017A&A...604A..17M,2022JGRA..12730011V}. This hypothesis is supported by the long-term stability of the light curves, maintained over $\sim$15 years, indicating a persistent and well-defined underlying geometry. Furthermore, as the putative companion orbits the target, it naturally induces changes in the longitude of the active field line ($\phi$) and in the half-opening angle of the emission cone ($\alpha$). Such variations are directly analogous to the Jupiter-Io system and provide a compelling explanation for the rotation-to-rotation variability observed in our lightcurves (Section ~\ref{subsubsec:alpha_var} and Section ~\ref{subsubsec:phi_var}). However, rather than the simplified scenario presented above, a more realistic interpretation might involve simultaneous changes in multiple physical parameters.

Although interaction between the target and a companion would provide one possible explanation for the presence of an active field line, there are other plausible scenarios which do not require a companion. For instance, the presence of corotation regions in the plasma torus where gas and dust accumulate, such as those reported in M-dwarfs by \cite{2024AJ....167...38B}, could create preferred sites for magnetospheric currents. In addition, complex and non-axisymmetric magnetic fields topologies have been proposed in late-M dwarfs \citep{2010MNRAS.407.2269M, 2015ApJ...802..106L}. Such magnetic configurations may create favored regions for particle acceleration and ECMI emission, and may even coexist with a large-scale axisymmetric dipolar field.
Likewise, purely geometrical effects, such as an anisotropic emission pattern or a non-axisymmetric hollow emission cone \citep[see][]{https://doi.org/10.1002/2015JA021038,2025ApJ...987....7G}, may also reproduce the apparent multiplicity of the L1 and L2 peaks. Future broadband and multi-epoch radio campaigns will be essential to disentangle these scenarios and determine whether the observed variability is driven by external modulation (e.g., companion), intrinsic magnetospheric dynamics or geometrical configurations.

However, independently of the origin of the active field line, we strongly favor the interpretation where both LCP peaks originate from different physical mechanisms. This conclusion is supported by these arguments; (i) the anticorrelated behavior of the brightest LCP (L1) and RCP peaks suggests a direct physical link between them, while the second LCP peak (L2) remains approximately constant; (ii) the stability of this anticorrelation over long timescales, which is also present in the observations reported by \cite{2008ApJ...684..644H} $\sim$15 years ago; (iii) the L2 LCP peak has no clear counterpart in the RCP emission; (iv) if both LCP peaks originate from the same geometrical emitting structure, such as the same auroral oval or active field line, changes in its geometry or emission conditions would be expected to affect both peaks in a correlated manner. The absence of an apparent correlated evolution between the two peaks is instead more consistent with their association with distinct emitting structures.

\subsection{Origin of the quiescent emission}
\label{subsec:Gyrosyn}

The VLA lightcurves shown in Figures ~\ref{fig:radio_lc} and ~\ref{fig:VLA_lc_ev} reveal a persistent emission component ocurring between the main peaks, remaining approximately constant throughout all VLA observations. This quiescent emission becomes more evident after subtracting the auroral model described in Section ~\ref{subsec:Auroral_Model} (see residuals in Figure ~\ref{fig:radio_fit_ev} and Figure ~\ref{fig:model_phi}). This component is detected in both LCP and RCP data, with flux densities of $\sim$50 -- 100~$\mu$Jy. In contrast, it is not detected in the VLBI data, as the mean rms ($\sim$150~$\mu$Jy) exceeds the expected flux of the quiescent component. 

The quiescent radio emission could originate from either gyrosynchrotron or synchrotron radiation. Resolved quiescent emission has previously been interpreted as synchrotron radiation from radiation belts surronding the M8.5 ultracool dwarf LSR J1835+3259 \citep{2023Sci...381.1120C, 2023Natur.619..272K}. In our observations, the emission remains unresolved and is confined within a region with a radius smaller than 52 $\rm R_{Jup}$ (see Section ~\ref{subsec:Radio_img}). Assuming a dipolar magnetic field and adopting the lower limit on the surface magnetic field of 4.3~kG, the size of the emitting region and the observing frequency do not provide strong constrains on the harmonic number of the emission, ($\rm~s < 1.5 \cdot 10^{5}$). This boundary is consistent with both gyrosynchrotron and synchrotron radiation \citep{2002ARA&A..40..217G}. Accordingly, we explore the implications of each one of both mechanisms in turn: (i) If the quiescent emission is produced by synchrotron radiation, the inferred magnetic field strength implies a characteristic Lorentz factor of $\gamma\lesssim280$, assuming that the electrons radiate predominantly near the critical synchrotron frequency and for nearly perpendicular pitch angles \citep{2002ARA&A..40..217G}. This corresponds to electron energies of $\rm E < 140$ MeV, somewhat higher than the tens of MeV inferred for Jupiter's radiation belts and for LSR~J1835+3259 \citep{2023Natur.619..272K, 2018JGRA..123.9110K}. These values should be regarded as upper limits because the emission may originate closer to the surface, where the magnetic field is stronger and lower-energy electrons would produce the observed emission. The spatial resolution of the data does not allow us to rule out this scenario. (ii) The quiescent emission of LSPM J0036+ was previously interpreted as gyrosynchrotron radiation by \cite{2017MNRAS.465.1995M}. In Appendix ~\ref{sec:Ap:gyro} we revisit this interpretation and show that our new observations remain consistent with that scenario.  A direct constraint on the extent of the emitting region or new linear polarization observations are required to establish whether the quiescent emission originates in a compact gyrosynchrotron source or in an extended synchrotron radiation belt.

\subsection{Searching for companions}
\label{subsec:Companion}

\cite{2010ApJ...715..724B} and \cite{2013ApJ...767..110P} reported the possible presence of a companion around LSPM J0036+1821 at an angular separation of $89.5 \pm 11.5$ mas. We used the VLBI observations to search for radio emission of the putative companion at this projected distance from the target, but did not detect any significant emission in any epoch. If such a companion exists and is radio-loud, we can place an upper limit on its emission of < 21~$\mu$Jy, corresponding to the 3$\sigma$ level of the VLBI epoch with the lowest rms noise (2021.45; Table ~\ref{tab:Obs}).

However, in the scenario where the active magnetic field line of LSPM J0036+1821 is driven by the presence of a companion, this proposed object could not be responsible for the observed interaction. At the distance proposed by \cite{2010ApJ...715..724B}, $0.78 \pm 0.1 $~AU, the companion would lie well beyond the sub-Alfv\'enic region, which is a requirement for an electrodynamic interaction analogous to the Jupiter-Io system. 

\section{Conclusions}
\label{sec:Con}

We report on 15 epochs of radio observations and 2 TESS sectors of the L3.5 dwarf LSPM J0036+1821 obtained over a time span of three years and covering different frequency bands. Our main findings can be summarized as follows:

\begin{enumerate}
\item We detect bursty auroral radio emission associated with the Electron Cyclotron Maser Instability (ECMI) mechanism at all epochs and instruments reported in Table ~\ref{tab:Obs}. In addtion, in the VLA observations, which provide higher sensitivity, we also detect a steady quiescent component.
\item We determine the rotation period of LSPM J0036+1821 to be P = $3.07908 \pm 0.00038$~h from optical observations and P = $3.07941 \pm 0.00020$~h from radio data, both of them are compatible within the 1$\sigma$ level and a precision of $\sim$1 s. The precise determination of both optical and radio periods places LSPM J0036+1821 among the few ultracool dwarfs for which the rotation period has been measured at both radio and optical wavelengths with a precision better than $\sim$1~min \citep[e.g.,][]{2014ApJ...788...23W, 2020Sci...368..169A}.
 \item We find the same peak morphology in the radio lightcurves in both VLA and VLBI averaged datasets, despite the observations spanning a period of three years.
 \item The auroral radio emission of LSPM J0036+1821 can be reproduced with a model consisting of two components: a main oval auroral encircling the magnetic poles and an active magnetic field line. This configuration closely resembles the auroral radio emission structures observed in the magnetospheres of Jupiter and Saturn.
 \item The rotation-to-rotation variability observed in the VLA lightcurves can be explained by small changes in either the half-opening angle of the ECMI emission cone or the longitude of the active field line. This effect reproduces the anticorrelated behavior observed between the RCP and LCP lightcurves. The stability of this anticorrelation evolution, which is also present in the observations reported by \cite{2008ApJ...684..644H}, suggests the presence of a stable mechanism modulating the morphology of the active field line. While an interaction with a companion, analogous to the Jupiter--Io system, represents one possible explanation, other physical scenarios may also account for the observed behavior. Additional observations are required to determine whether the variability exhibits a periodic pattern and to constrain its physical origin.
 \item The quiescent emission detected in the VLA observations is compatible with both gyrosynchrotron and synchrotron emission. Our measurements are consistent with the gyrosynchrotron interpretation proposed by \cite{2017MNRAS.465.1995M}; however, the unresolved nature of the source prevents us from uniquely identifying the emission mechanism.
 \item We do not detect radio emission from the candidate companion to LSPM J0036+1821 reported at a projected separation of  $\sim90$ mas. If such a companion exists, we place an upper limit on its radio flux of < 21~$\mu$Jy.

 \end{enumerate}

LSPM J0036+1821 has long been recognized as one of the benchmark radio-emitting ultracool dwarfs, and the results presented here further reinforce its importance as a laboratory for magnetospheric studies. The long-term stability of the emission over $\sim$15 years, combined with the short-term variability that can be explained by changes in the emission cone geometry and the longitude of the active field line, is compatible with the presence of an external modulation mechanism, such as a companion driving the emission. However, alternative scenarios cannot be ruled out and should be explored with more comprehensive models and future observations. In this context, the ExPRES code \citep{2019GeoRL..4611606L} offers a promising tool for future work, as it enables geometrical simulations with a self--consistent calculation of the full beaming pattern. Regardless of the underlying mechanism, LSPM J0036+1821 remains a particularly promising target for constraining the physical processes responsible for auroral radio emission in the ultracool dwarf regime.

\section*{Data availability}

The VLA and VLBA data used in this work are publicly available through the NRAO Science Data Archive (\url{https://data.nrao.edu/portal/}) under project codes SR0228 for the VLA and BC250 and BC267 for the VLBA. The EVN observations are available through the EVN Data Archive (\url{https://evn-vo.jive.nl/}), provided by JIVE on behalf of the European VLBI Network, under project codes EC067C and EC077C. The TESS observations from Sectors 17 and 57 are publicly available from the Mikulski Archive for Space Telescopes (MAST; \url{https://archive.stsci.edu/missions-and-data/tess}). The auroral emission modelling was performed using the \texttt{magnetic-geometry-retrieval} code described by \citet{2024A&A...692A..66K} and \citet{2024A&A...682A.170B}, available at \url{https://github.com/robkavanagh/magnetic-geometry-retrieval/}, which makes use of \texttt{UltraNest} (\url{https://johannesbuchner.github.io/UltraNest/}). The gyrosynchrotron emission was modeled using the FAST gyrosynchrotron codes presented in \cite{2021ApJ...922..103K}, available in \url{https://doi.org/10.5281/zenodo.5139156}.

\begin{acknowledgements}

D.M.C, J.B.C, J.B. and J.C.G. were supported by MCIN/AEI through the project PID2023-147883NB-C22, and by the Generalitat Valenciana through the Prometeo Programme for Research Excellence (project CIPROM/2022/64). P.A.M.P acknowledges use of the grants RYC2021-031173-I and PID2022-137241NB-C42 funded by MCIN/AEI/10.13039/501100011033 and by the 'European UNion NextGenerationEU/PRTR'. JMV and CO’T acknowledge support from a Royal Society - Research Ireland University Research Fellowship (URF/1/221932, RF/ERE/221108). The European VLBI Network is a joint facility of independent European, African, Asian, and North American radio astronomy institutes. Scientific results from data presented in this publication are derived from observations made with the European VLBI Network. This research has made use of archival data from the Karl G. Jansky Very Large Array (VLA) and the Very Long Baseline Array (VLBA), accessed through the NRAO Science Data Archive. The National Radio Astronomy Observatory is a facility of the National Science Foundation operated under cooperative agreement by Associated Universities, Inc.
      
\end{acknowledgements}

\bibliographystyle{aa} 
\bibliography{lib}

\begin{appendix}

\onecolumn

\section{Radio observations campaigns}

\begin{table*}[h!]
\centering
\caption {Radio observations campaigns of LSPM J0036+1821.}
\label{tab:Obs}
\begin{tabular}{
c c 
S[table-format=2]
@{--}
c
@{--}
S[table-format=4]
c
c
@{:}
c
@{--}
c
@{:}
c
S[table-format=3.1]
@{$\times$}
S[table-format=3.1]
@{,\,}
S[table-format=3.2]
c
@{$\pm$}
c
c
@{$\pm$}
c}
\toprule
Antennas & Project 
& \multicolumn{4}{c}{Observing Date}
& \multicolumn{4}{c}{UT Range} 
& \multicolumn{3}{c}{Beam}
& \multicolumn{2}{c}{{$F_{{\rm peak}, I}$}}
& \multicolumn{2}{c}{{$F_{{\rm peak}, V}$}} \\

&  
& \multicolumn{4}{c}{}
& \multicolumn{4}{c}{}
& \multicolumn{3}{c}{(mas $\times$ mas, deg)}
& \multicolumn{2}{c}{($\mu$Jy/beam)}
& \multicolumn{2}{c}{($\mu$Jy/beam)} \\
\midrule

EVN  & EC067C  & 03 & Mar & 2019 & (2019.17) & 08 & 30 & 10 & 59 
& 0.9  & 2.8  & 9.02   & 78  & 16  & 50  & 15 \\

EVN  & EC077C  &  15 & Jun & 2021 & (2021.45) & 03 & 30 & 08 & 22 
& 2.7  & 2.9  & -76.33 & 165 & 7   & -130 & 8 \\

EVLA & SR0228  & 07 & Dec & 2021 & (2021.93) & 00 & 20 & 09 & 00 
& 770.0 & 730.0 & 79.90 & 160 & 4   & -130 & 4 \\

VLBA & BC250B1 &   01 & Mar & 2020 & (2020.17) & 18 & 58 & 22 & 35 
& 1.7  & 3.8  & 6.32   & 274 & 20  & -99  & 21 \\

VLBA & BC250B2 &   23 & Mar & 2020 & (2020.23) & 17 & 32 & 21 & 09 
& 1.7  & 3.8  & 4.80   & 307 & 32  & -154 & 35 \\

VLBA & BC267B1 &   14 & Feb & 2021 & (2021.12) & 20 & 05 & 23 & 01 
& 1.5  & 3.8  & 6.03   & 111 & 23 & 60   & 23 \\

VLBA & BC267B2 &   26 & Mar & 2021 & (2021.23) & 17 & 29 & 20 & 24 
& 1.8  & 4.0  & 1.90   & 202 & 17  & -82  & 16 \\

VLBA & BC267B3 &   11 & Apr & 2021 & (2021.28) & 16 & 25 & 19 & 22 
& 1.8  & 4.0  & 3.90   & 203 & 16  & -118 & 16\\

VLBA & BC267B4 &   02 & May & 2021 & (2021.33) & 14 & 57 & 17 & 53 
& 1.6  & 3.8  & 4.07   & 248 & 15  & -102 & 15 \\

VLBA & BC267B5 &   07 & May & 2021 & (2021.35) & 14 & 53 & 17 & 50 
& 1.8  & 4.2  & -3.80  & 140 & 18  & -80  & 17 \\

VLBA & BC267B6 &   21 & Jun & 2021 & (2021.47) & 11 & 47 & 14 & 43 
& 1.6  & 3.8  & 2.40   & 133 & 15  & -55 & 15 \\

VLBA & BC267B7 &   29 & Aug & 2021 & (2021.66) & 07 & 15 & 10 & 20 
& 1.3  & 3.9  & -5.40  & 99 & 20  & \multicolumn{2}{c}{$<46$} \\

VLBA & BC267B8 &   11 & Oct & 2021 & (2021.78) & 04 & 26 & 07 & 22 
& 1.7  & 5.2  & -13.70 & 215 & 22  & -110 & 22 \\

VLBA & BC267B9 &   18 & Dec & 2021 & (2021.96) & 23 & 11 & 03 & 12 
& 1.5  & 4.0  & 6.30   & 146 & 17  & -70  & 17 \\

\bottomrule
\end{tabular}
\end{table*}

\section{VLA lightcurves}

   \begin{figure*}[h!]
        \centering
        \includegraphics[width=15.25cm]{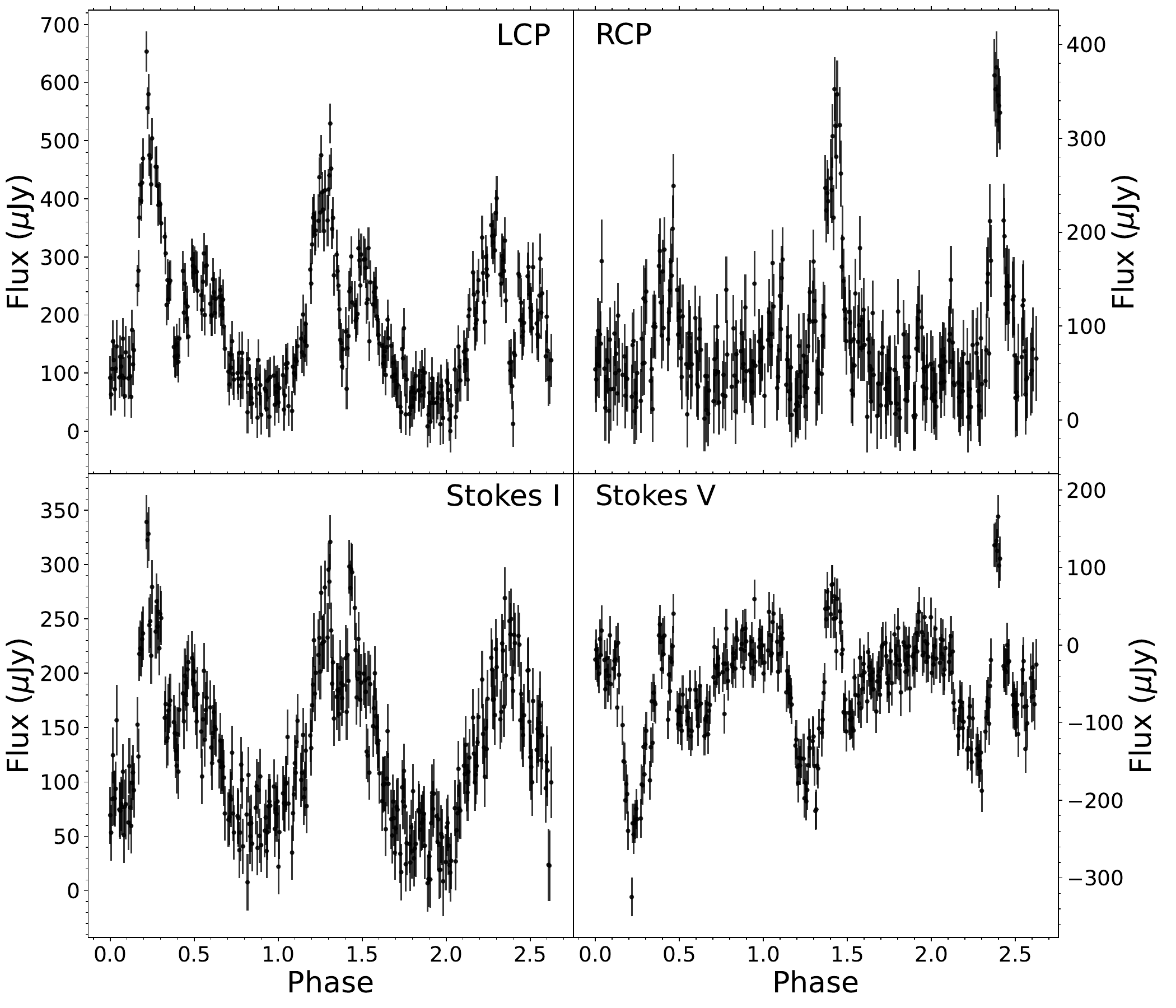}
        \caption{VLA lightcurves. Top panel shows LCP (left) and RCP (right). Bottom panels show Stokes I (left) and Stokes V (right). Black data points represent VLA flux densities averaged every 60 s.}
        \label{fig:VLA_lc_ev}%
    \end{figure*}
\clearpage

\twocolumn

\section{Thickness of the ECMI hollow cone emission}

\label{sec:Ap:model}

The active-field-line model of \citet{2024A&A...692A..66K} computes the angle $\gamma$ between the observer's line of sight and the local magnetic-field direction along a dipolar active field line. ECMI emission is brightest when $\gamma$ equals the cone half-opening angle $\alpha$. In the original AFL prescription, the flux from the northern and southern magnetic hemispheres is combined as a signed Stokes-$V$ light curve,
\begin{equation}
F_V = F_B \left[
\exp\left(-\frac{1}{2}\left[\frac{\gamma_N-\alpha}{\Delta\alpha}\right]^2\right)
-
\exp\left(-\frac{1}{2}\left[\frac{\gamma_S-\alpha}{\Delta\alpha}\right]^2\right)
\right],
\end{equation}
where $\Delta\alpha$ is used as the Gaussian width of the beaming profile. In that form, the Gaussian wings extend to all angular offsets. Therefore, although $\Delta\alpha$ sets the scale over which the flux decreases, it is not a strict full cone-wall thickness. Under this prescription, sufficiently large peak flux values may produce detectable emission at angular distances significantly larger than $\Delta\alpha$ from the center of the emission cone.

For the present work we keep the geometrical calculation of $\gamma_N$ and $\gamma_S$ from \citet{2024A&A...692A..66K}, but modify the emissivity prescription so that $\Delta\alpha$ can be interpreted as the physical full thickness of the hollow-cone wall. For each hemisphere $H\in\{N,S\}$, we define a visibility window
\begin{equation}
W_H =
\begin{cases}
1, & |\gamma_H-\alpha| \leq \Delta\alpha/2,\\
0, & |\gamma_H-\alpha| > \Delta\alpha/2,
\end{cases}
\end{equation}
and an emissivity profile
\begin{equation}
G_H = W_H \,F \left[\,\exp\left(-\frac{1}{2}
\left[\frac{\gamma_H-\alpha}{\sigma}\right]^2\right)\right],
\qquad
\sigma = \frac{\Delta\alpha}{\rho}.
\end{equation}
The additional dimensionless parameter $\rho$ controls the steepness of the profile inside the allowed cone wall. At the edge of the wall, $|\gamma_H-\alpha|=\Delta\alpha/2$, the relative flux is $\exp(-\rho^2/8)$. This choice ensures that the fitted $\Delta\alpha$ corresponds to an effective full wall thickness, while still allowing a smooth profile within the wall.

We also fit the two circular-polarization light curves and the Stokes I and V parameters directly. Assuming x-mode emission and the IAU Stokes-$V$ convention, the northern hemisphere contributes to RCP and the southern hemisphere contributes to LCP.

The main-oval component is implemented analogously. We assume that the ECMI emission cones are uniformly distributed around the magnetic poles. Each element has the same beaming prescription as above, with the main-oval parameters $(\theta_{\rm MO},\alpha_{\rm MO},\Delta\alpha_{\rm MO},F_{\rm MO})$. The contribution of each element is evaluated independently and then summed over the ring. The flux normalization per element is $F_{\rm MO}/N_{\rm ring}$.

\onecolumn

\section{Corner plot of the phase-folded radio lightcurves}

   \begin{figure}[h!]
   \centering
   \includegraphics[width=15cm]{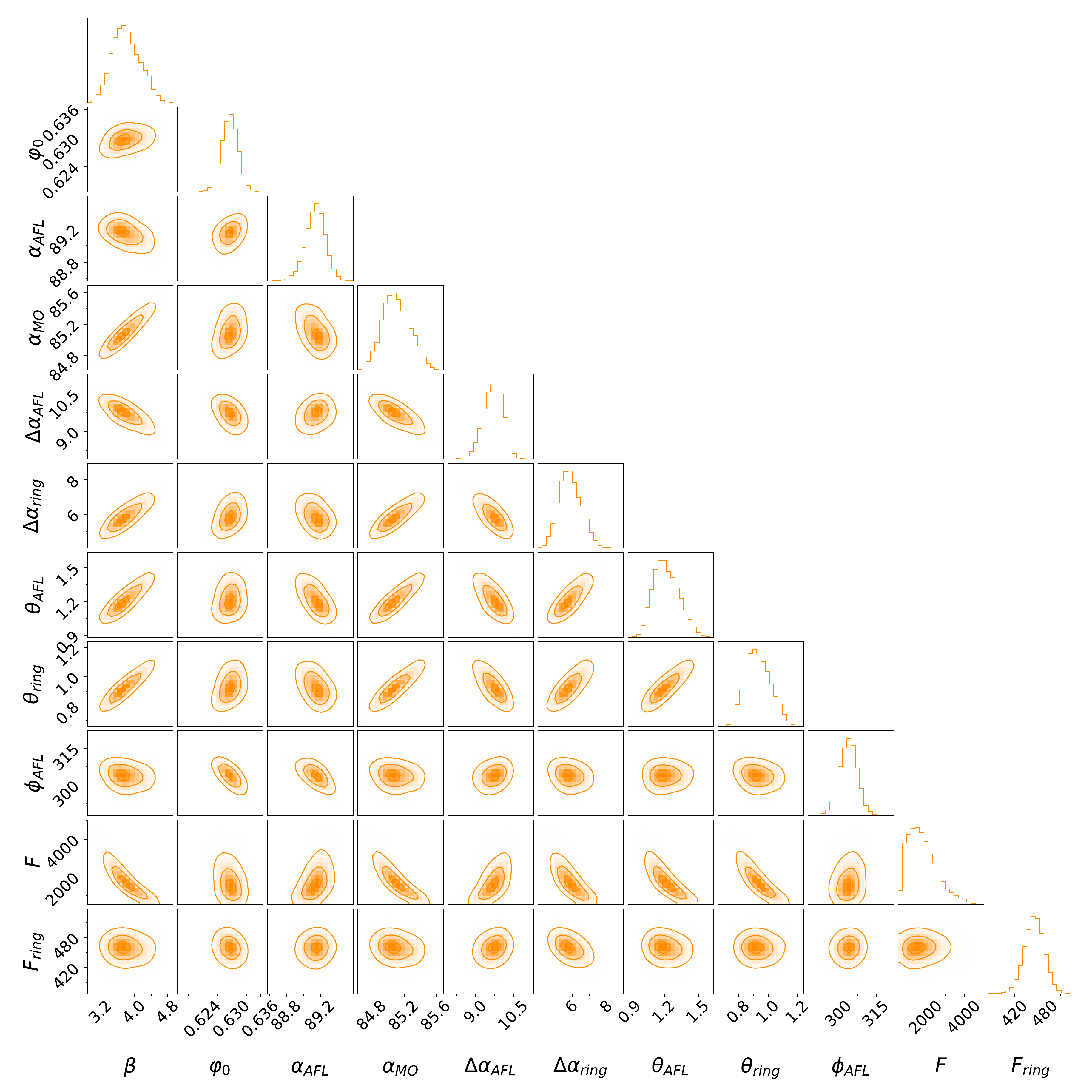}
      \caption{Corner plot showing the posterior distribution of the MCMC fiting. Diagonal panels show posterior distribution of the model for the parameters shown in Table ~\ref{table:AURO_MODEL}. The rest of the panels represent the covariance between the parameters, darker colors represent higher probability. }
         \label{fig:CP_MEANMODEL}
   \end{figure}

\clearpage

\section{Model fit free $\phi$}

   \begin{figure*}[h!]
        \centering
        \includegraphics[width=16cm]{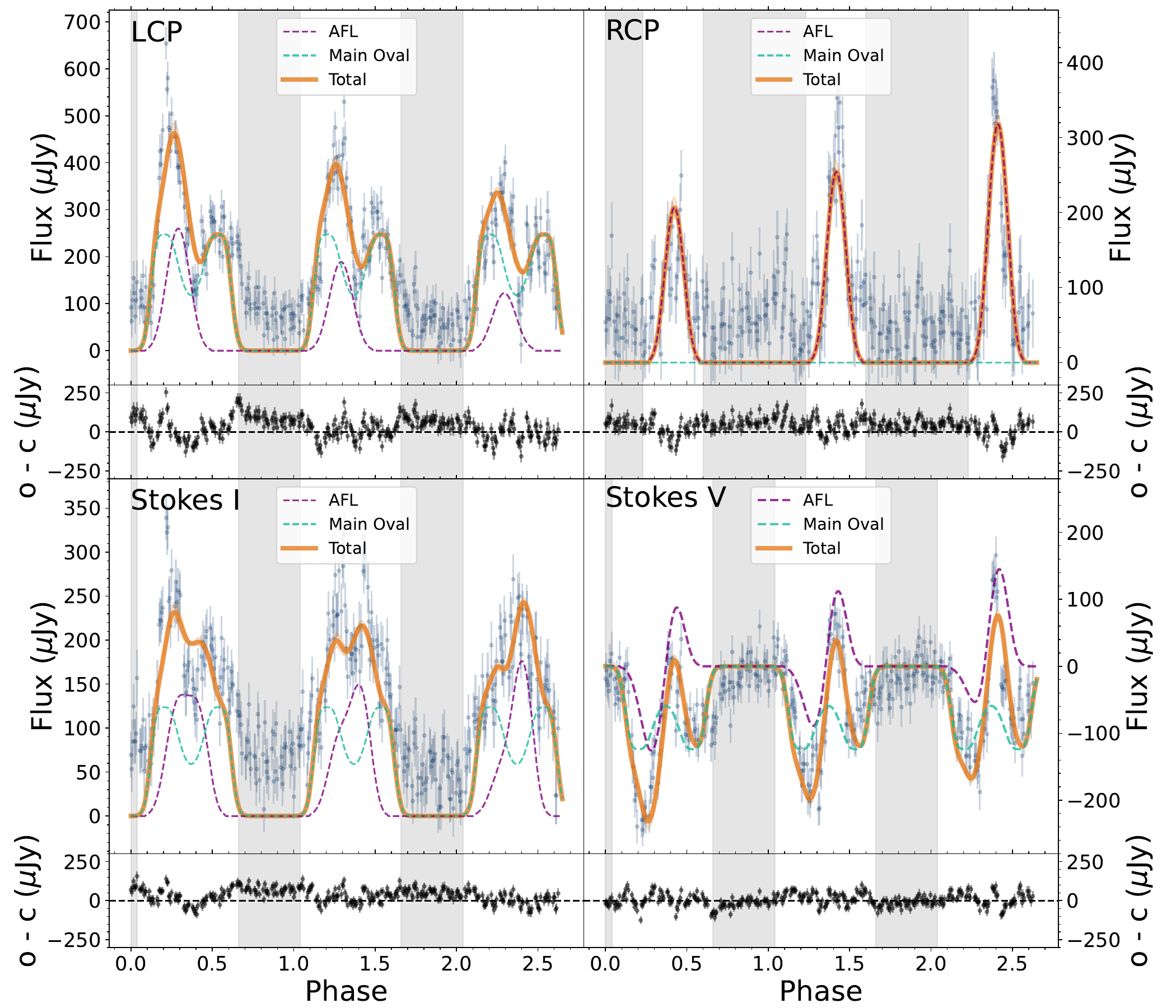}
        \caption{Same as Figure ~\ref{fig:radio_fit_ev} but leaving $\phi$ as free parameters between rotations. The fixed parameters are listed in Table ~\ref{table:AURO_MODEL} and Table ~\ref{table:AURO_MODEL_ROT_PHI} shows the prior and posterior distribution of $\varphi$ which is changing from one rotation to the next.}
        \label{fig:model_phi}%
    \end{figure*}

\twocolumn
\section{Quiescent emission: Gyrosynchrotron}
\label{sec:Ap:gyro}

\begin{figure}[h!]
        \centering
        \includegraphics[width=7.75cm]{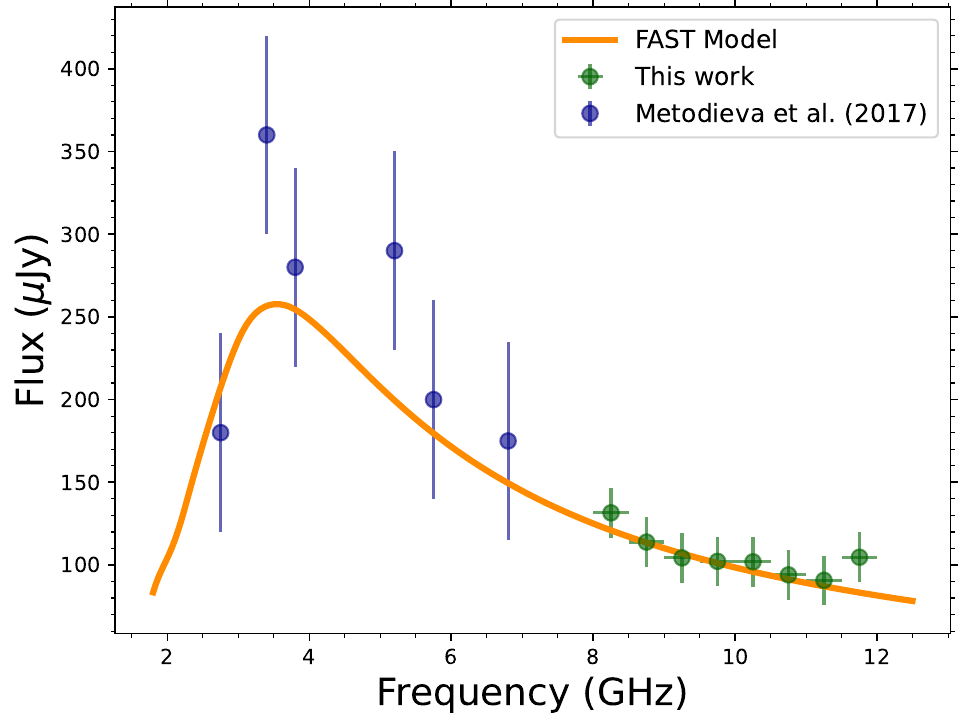}
        \caption{Radio spectra of LSPM J0036+1821 between 2 -- 12~GHz. Blue dots correspond to the measurements reported by \cite{2017MNRAS.465.1995M}, while green dots represent the measurements from this work. The gyrosynchrotron model derived by \cite{2017MNRAS.465.1995M} is shown in orange. The model is consistent with the new data points presented in this work.}
        \label{fig:gyro_em}%
    \end{figure}

\cite{2017MNRAS.465.1995M} reported a quiescent emission component in LSPM J0036+1821 between 2 and 7~GHz attributed to the gyrosynchrotron mechanism and modeled their data using the {\sc FAST} gyrosynchrotron code \citep{2021ApJ...922..103K}. In Section ~\ref{subsec: Radio_em}, we derived a lower limit for the magnetic field strength of $B > 4.3$~kG, corresponding to the upper end of the values reported by \cite{2017MNRAS.465.1995M} (see their Table 5). We built a radio spectrum complementing Metodieva's data with flux density values derived from VLA images of LSPM J0036+1821 obtained separately from adjacent 0.5~GHz bandwidth data sets. To account only for the quiescent emission, we used the time range corresponding to the gray shaded areas indicated in the Figure ~\ref{fig:radio_fit_ev}. Following \cite{2017MNRAS.465.1995M}, we adopted the same set of parameters in the FAST gyrosynchrotron model and overlaid the resulting model on our data. The results are presented in Figure ~\ref{fig:gyro_em}. We find that our observations are consistent with the fit obtained by \cite{2017MNRAS.465.1995M}. This solution implies an emitting region with a size of $\sim$4 Jupiter radii, a local magnetic field strength of $\sim$200~G, a moderate electron energy spectral index of $\delta \sim$2.4 and a viewing angle of $\theta \sim$140º. However, we can not unambiguously determine gyrosynchrotron mechanism to be responsible of the emission as the extend of the source does not allow us to a detailed characterization of the emission.

\onecolumn

\clearpage

\end{appendix}
\end{document}